\documentclass[aps,pre,a4paper,twocolumn,preprintnumbers]{revtex4}

\usepackage[utf8]{inputenc}
\usepackage{graphicx}
\usepackage{psfrag}
\usepackage{color}
\usepackage{amsmath}
\usepackage{amscd}
\usepackage{amssymb}
\usepackage{amsthm}
\usepackage{caption}
\usepackage{ragged2e}
\usepackage{nicefrac}
\usepackage{multirow}
\usepackage[caption=false]{subfig}
\usepackage{pslatex}

\DeclareCaptionJustification{reallyjustified}{\justifying}
\begin{document}

\title{Mesoscopic approach to subcritical fatigue crack growth}

\author{Maycon S. Araújo} \email[]{maycon@if.usp.br}
\affiliation{Departamento de Física Geral, Instituto de Física, Universidade de 
São Paulo \\ Caixa Postal 66318, 05314-970, São Paulo, SP, Brazil}

\author{André P. Vieira} \email[]{apvieira@if.usp.br}
\affiliation{Departamento de Física Geral, Instituto de Física, Universidade de 
São Paulo \\ Caixa Postal 66318, 05314-970, São Paulo, SP, Brazil}

\author{José S. Andrade Jr.} \email[]{soares@fisica.ufc.br}
\affiliation{Departamento de Física, Universidade Federal do Ceará \\ Caixa 
Postal 6030, 60451-970, Fortaleza, CE, Brazil}

\author{Hans J. Herrmann} \email[]{hans@ifb.baug.ethz.ch}
\affiliation{Departamento de Física, Universidade Federal do Ceará \\ Caixa 
Postal 6030, 60451-970, Fortaleza, CE, Brazil}
\affiliation{Computational Physics, Institut für Baustoffe (IfB), 
ETH-Hönggerberg \\ Schafmattstrasse 6, HIF E 12, CH-8093, Zürich, Switzerland}

\date{\today}


\begin{abstract}
We investigate a model for fatigue crack growth in which 
damage accumulation is assumed to follow a power law of the local
stress amplitude, a form which can be generically justified on
the grounds of the approximately self-similar aspect of microcrack 
distributions. Our aim is to determine the relation between model
ingredients and the Paris exponent governing subcritical crack-growth
dynamics at the macroscopic scale, starting from a single small
notch propagating along a fixed line. By a series of analytical 
and numerical calculations, we show that, in the absence of 
disorder, there is a critical damage-accumulation exponent $\gamma$,
namely $\gamma_c=2$, separating two distinct regimes of behavior
for the Paris exponent $m$. For $\gamma>\gamma_c$, the Paris exponent
is shown to assume the value $m=\gamma$, a result which proves robust
against the separate introduction of various modifying ingredients. 
Explicitly, we deal here with (i) the requirement of a minimum stress
for damage to occur; (ii) the presence of disorder in local 
damage thresholds; (iii) the possibility of crack healing.
On the other hand, in the regime $\gamma<\gamma_c$ the Paris exponent
is seen to be sensitive to the different ingredients added to the model,
with rapid healing or a high minimum stress for damage leading to 
$m=2$ for all $\gamma<\gamma_c$,
in contrast with the linear dependence $m=6-2\gamma$ observed for very
long characteristic healing times in the absence of a minimum stress
for damage. Upon the introduction of disorder on the local fatigue thresholds,
which leads to the possible appearance of multiple cracks along the
propagation line, the Paris exponent tends to $m\approx 4$ for
$\gamma\lesssim 2$, while retaining the behavior $m=\gamma$ for $\gamma\gtrsim 4$.

\end{abstract}

\maketitle


\section{Introduction}

Fracture phenomena are quite common in nature and play a fundamental role in many 
situations of interest for science and technological applications 
\cite{Suresh1998,Donald-Turcotte-Book}. 
%
Despite many advances in materials science and applied mechanics along the past 
decades, the full description of such problems remains a great challenge to 
physicists 
and engineers \cite{Vladimir-Bolotin-Book}. However, it is a well-known fact 
that the presence of cracks within a material can magnify by several times the 
effect of the external stresses applied, causing a strong reduction in its 
strength and inducing rupture at a stress very much lower than that needed to 
break the atomic bonds in a flawless, regular arrangement 
\cite{Vladimir-Bolotin-Book,Marder-Book}.

Scaling arguments developed by Griffith \cite{Griffith1921} show that a single 
crack, after reaching some critical length, will propagate spontaneously within 
the material, causing its catastrophic failure. Below that critical length, 
many 
kinds of external mechanisms occurring on relatively slow time scales can 
dominate the crack dynamics, defining a subcritical regime of crack growth
\cite{Suresh1998}. Among those mechanisms we highlight the occurrence of 
fatigue as the result of a progressive accumulation of damage throughout the 
material when submitted to cyclic load
\cite{Suresh1998,Vladimir-Bolotin-Book,Krupp-Book}.

In general, subcritical fatigue crack propagation is well described by an 
empirical law largely used in engineering practice, known as the Paris
(or Paris-Erdogan) law 
\cite{Paris1963}, which states that the growth rate of a linear crack under 
cyclic load follows a power law of the stress-intensity factor,
with an exponent $m$,
\begin{align}
\frac{da}{dN}=C(\Delta K)^{m}\sim a^{m/2}.\label{paris}
\end{align}
Here $a$ is the crack half-length, $N$ is the number of loading cycles
applied to the material, $da/dN$ is the crack growth rate (proportional to the 
crack tip speed), $\Delta K\equiv g\Delta\sigma_0\sqrt{\pi a}$ is 
the amplitude of the stress-intensity factor of the crack,  
$\Delta\sigma_0$ and $g$ being the stress amplitude
and a geometrical factor, respectively, while $m$ (the Paris exponent) and $C$
are parameters which may depend on both the material properties
and the experimental conditions. 
Numerous experiments confirm the validity of this law 
over several orders of magnitude for a wide variety of materials and loading
conditions \cite{Suresh1998,Vladimir-Bolotin-Book}.

Despite its simplicity and practical importance, 
a systematic understanding of this law on 
physical grounds is still lacking, especially as regards the 
determination of an explicit relation between the Paris exponent $m$ and 
microscopic parameters of a given material.
An intermediate step was taken by three of the authors of the present 
paper \cite{Vieira2008}, who were able to show that
the Paris law indeed emerges from a damage-accumulation rule defined
by a power law of the external stress amplitude, with a characteristic 
exponent $\gamma$, whose relation with the Paris exponent $m$ can
be determined via a combination of analytical and numerical
calculations. Although such a damage-accumulation 
rule can be justified by invoking self-similarity concepts
\cite{Botvina1986}, a first-principle calculation of the damage-accumulation exponent
$\gamma$ for a given material remains challenging.
Nevertheless, assuming such a damage-accumulation rule on phenomenological grounds,
it is possible to show \cite{Vieira2008} that, in the absence of 
disorder, there is a critical damage-accumulation exponent $\gamma$,
namely $\gamma_c=2$, separating two distinct regimes of behavior
for the Paris exponent $m$. For $\gamma>\gamma_c$, the Paris exponent
assumes the value $m=\gamma$, while for $\gamma<\gamma_c$ a different
linear relation, $m=6-2\gamma$, is verified.

Our aim in this paper is to further explore the consequences of the
dynamics associated with a power-law damage accumulation rule,
both in the uniform limit and 
in combination with disorder in the local rupture thresholds.
Regarding disorder, some progress has already been made in Ref. 
\cite{Oliveira2012} by a mapping to a random-fuse problem,
which was solved numerically. Here we combine results from
linear-elastic fracture mechanics with an independent-crack
approximation to perform a thorough study of the effects of disorder
on the relation between the damage-accumulation exponent $\gamma$
and the Paris exponent $m$. We also investigate the effects of 
introducing a healing mechanism which lowers the local damage
throughout the material as time passes.
We present evidence that the relation $m=\gamma$ for $\gamma>\gamma_c$ 
is robust against the separate introduction of various modifying ingredients, 
but that in the regime $\gamma<\gamma_c$ the Paris exponent
is sensitive to the different ingredients added to the model,
with rapid healing or a high minimum stress for damage leading to 
$m=2$ for all $\gamma<\gamma_c$, while disorder leads to
$m\approx 4$.

The paper is organized as follows. The basic ingredients
of the model are presented in Sec. \ref{model}, with
the next two sections dedicated to investigating
the uniform limit in the absence of healing. 
The behavior of the model in the presence of disordered
local damage thresholds in discussed in Sec. \ref{sec:disorder}.
Healing effects in the uniform limit are introduced and
discussed in Sec. \ref{sec:healing}. The final section
summarizes our findings.


\section{The basic model \label{model}}

\begin{figure}
\centering
\includegraphics[width=\columnwidth]{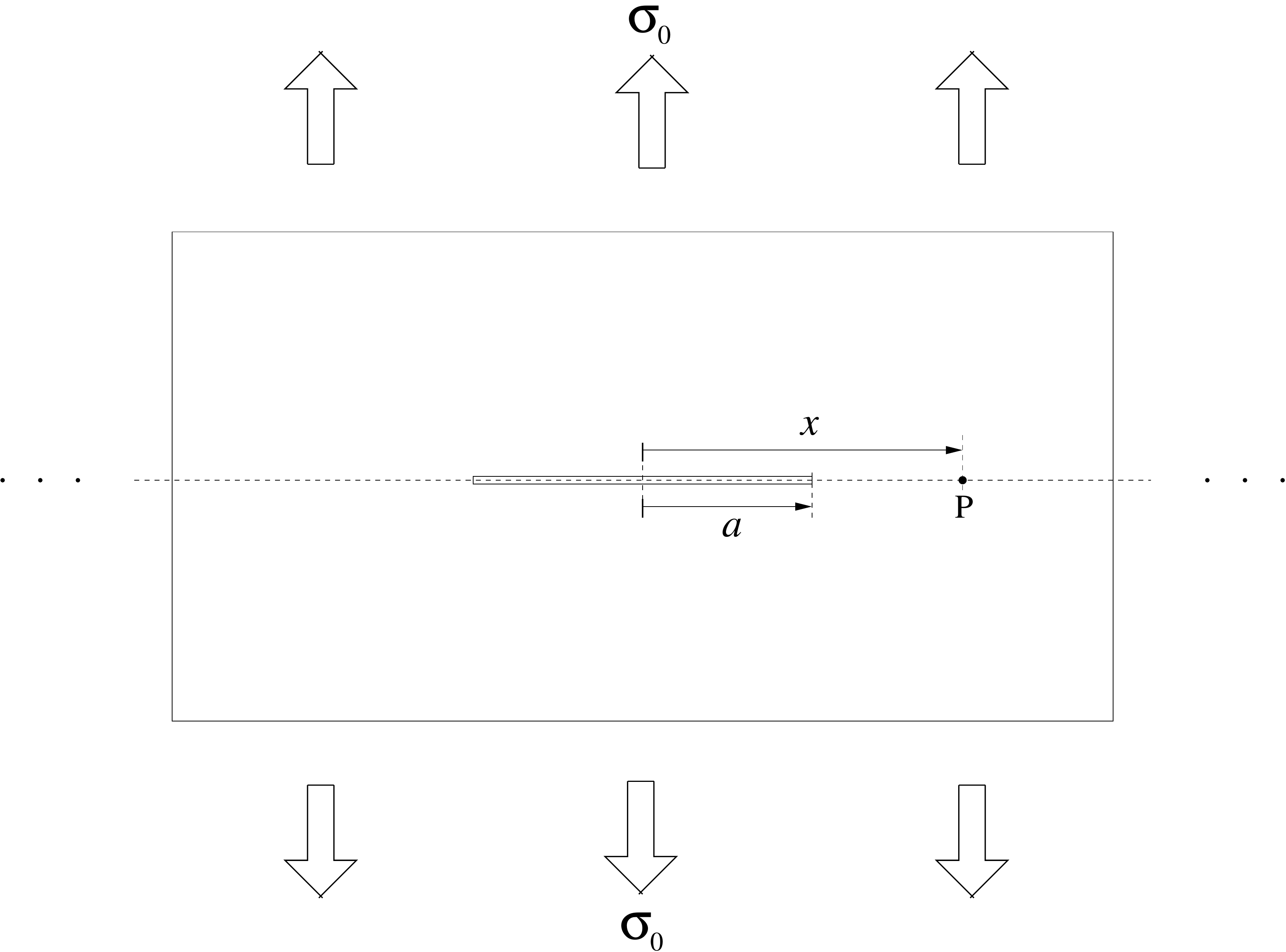}
\caption{A very thin elliptical crack of half-length $a$ 
propagating along the direction of its major axis in 
a two-dimensional sample of material subject to an external stress $\sigma_0$.
The crack propagation line is indicated by the dashed 
line, and $x$ is the coordinate of a given point $P$ relative to the midpoint
of the crack.}
\label{fig1}
\end{figure}

\begin{figure}
\centering
\includegraphics[width=\columnwidth]{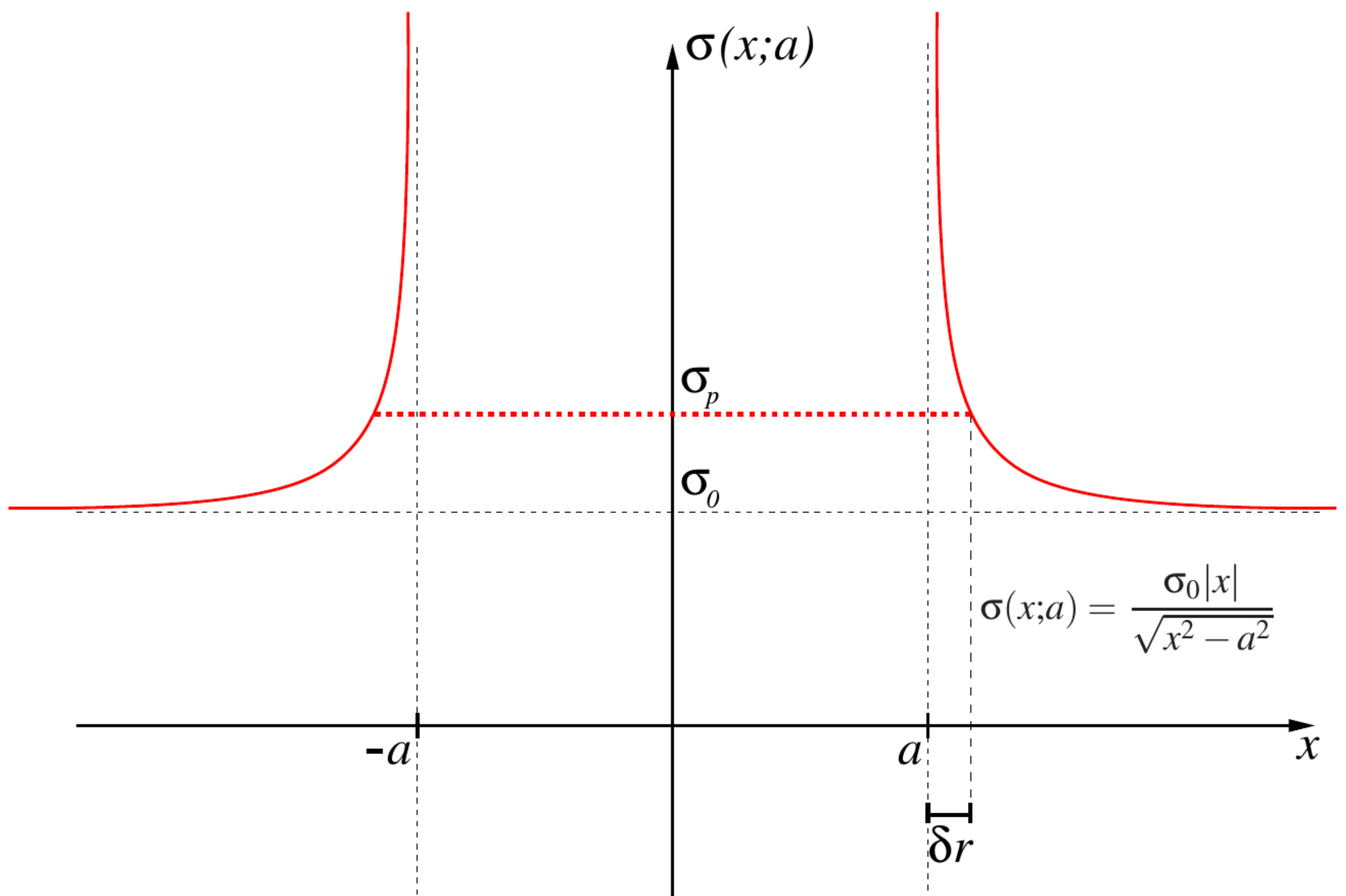}
\caption{(Color online) Sketch of the stress field $\sigma(x;a)$ along 
the propagation line of the crack, whose midpoint is at
$x_0=0$. Except for the presence of the crack, the medium is homogeneous. 
The cutoff value $\sigma_p$ stands for the 
stress attributed to plastic effects within a zone around the crack tip, 
whose linear dimension is assumed here to be smaller than $\delta r$.}
\label{fig2}
\end{figure}

In the present section we define the model and discuss schematically the 
dynamics of crack growth. The next sections deal with particular cases
and extensions.

Following Ref. \cite{Vieira2008},
we assume that a single thin elliptic crack is initially produced
in an infinite two-dimensional sample of a linear-elastic material.
The sample is subject to cyclic loading, with an external stress
$\sigma_0$ transverse to the major axis of the crack. We further assume that
the crack grows only along its major axis, so that crack propagation
becomes essentially a one-dimensional problem, as shown in Fig. \ref{fig1}.

Along the crack line, we discretize space so that the crack grows by 
the rupture of elements of fixed length $\delta r$, and assume that,
when the crack has length $2a$,
the element at position $x$ experiences a stress given by
$\sigma(x+\delta r; a)$. This assumption prevents the appearance of 
divergences in the stress field around the crack tip and, to a first approximation, is 
consistent with the fact that linear-elasticity theory must break down in the immediate 
vicinity of the crack tip, giving rise to a fracture process zone or
plastic zone \cite{Suresh1998,Alava2006}. We assume in this work that
the size of the fracture process zone is smaller than the discretization length $\delta r$.
We also assume that
the relaxation time of the material is much shorter than the period of the
loading cycle, so that crack propagation can be investigated within a quasistatic 
approximation, according to which the system always reaches its equilibrium 
state between two successive crack-growth events.

In the continuum limit, and within linear-elasticity theory, 
the local stress $\sigma(x;a)$ along the crack line is 
given by \cite{Marder-Book}
\begin{equation}
\sigma(x;a)=\sigma_0\displaystyle\frac{|x-x_0|}{\sqrt{\left(x-x_0\right)^{2}-a^{2}}} 
\label{stress-range}, 
\end{equation}
where $\sigma_0$ is the external stress applied to 
the material, $x$ is the coordinate of the point of interest,
$x_0$ is the coordinate of the midpoint of the crack
and $2a$ is the crack length (see Fig. \ref{fig2}).

\begin{figure}
\centering
\includegraphics[width=\columnwidth]{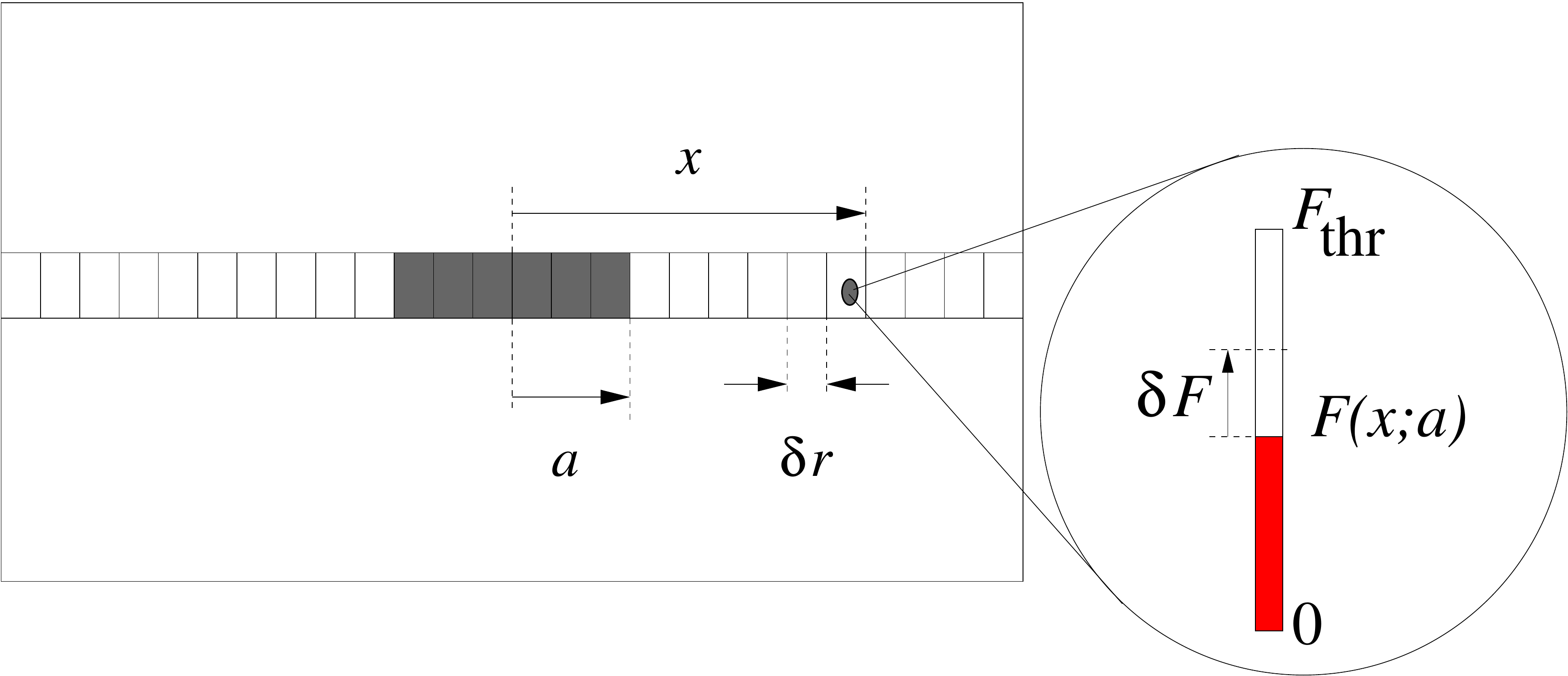}
\caption{(Color online) Schematic diagram representing the damage-accumulation 
process for a given element at position $x$ in a configuration 
with crack half-length $a$. The damage accumulated $F(x;a)$ is depicted by 
a life-bar with level labeled in red which can receive a damage increment $\delta 
F$ until it reaches the damage-accumulation threshold $F_{\mathrm{thr}}$.
}
\label{fig3}
\end{figure} 

Sufficiently close to the crack tips, we 
obtain an asymptotic expression for $\sigma(x;a)$,
\begin{equation}
\sigma(x;a)\simeq\frac{K}{\sqrt{2\pi(\left|x-x_0\right|-a)}}
\label{stress-field},
\end{equation}
defining the stress intensity factor $K=\sigma_0\sqrt{\pi a}$ for 
this particular geometry. 

We postulate that cyclic loading with an external stress amplitude 
$\Delta\sigma_0\equiv \sigma_{0,\mathrm{max}}-\sigma_{0,\mathrm{min}}$
leads to fatigue damage accumulation in each
element along the crack line according to the rule
\begin{equation}
\delta F(x;a)=f_0\delta t(a)[\Delta\sigma(x;a)]^\gamma,\label{damage}
\end{equation}
where $\delta F(x;a)$ is the damage increment in the element 
located at position $x$ during the time interval $\delta t(a)$ when
the crack remained with length $2a$, $\Delta\sigma(x;a)$ is the
corresponding local stress amplitude,
$\gamma$ is a phenomenological
damage accumulation exponent and $f_0$ is a constant setting the
time scale, being proportional to the inverse duration of
the loading cycle; see Fig. \ref{fig3} for an illustration. 

Therefore, the damage at position $x$ when
the crack is about to grow from length $2a$ is given by the
relation
\begin{equation}
F(x;a)=F(x;a^\prime)+\delta F(x;a),
\label{recurrence}
\end{equation}
in which $2a^\prime$ is the previous crack length. When the crack always
advances symmetrically with respect to the midpoint of the initial crack,
we have $a^\prime = a-\delta r$.

A heuristic motivation for the power-law dependence of the damage increment 
can be formulated by invoking concepts of self similarity and 
fractality 
commonly observed in spatial patterns related to crack propagation and 
fragmentation processes 
\cite{Botvina1986,Krajcinovic-Book,Herrmann-Book,Herrmann1989,Herrmann1989-2,Herrmann1991-2},
and assuming that the most important contribution to damage accumulation comes
from the local stress amplitude.

Finally, we assume that an element at position $x$ ruptures when the 
corresponding accumulated damage reaches a threshold $F_\mathrm{thr}(x)$.
In the uniform limit, $F_\mathrm{thr}(x)\equiv F_\mathrm{thr}$ for all $x$,
elements break sequentially, starting from the initial crack tips, and the crack
advances symmetrically. In the general case, as shown below, elements
far from the crack tip can suffer early rupture, leading to irregular crack growth
and to the presence of multiple cracks.
In all cases, we focus on the growth of the initial crack --- or 
\emph{main} crack --- which may involve secondary 
cracks when these coalesce with the main crack.

The main crack advances when the accumulated damage in one or both 
elements at the crack tips reaches the corresponding threshold. 
Equations (\ref{damage}) and (\ref{recurrence}) allow the
calculation of the number of cycles since the last growth event
and of the updated accumulated damage along the crack line.

\begin{figure}
\centering
\subfloat{
\includegraphics[width=\columnwidth]{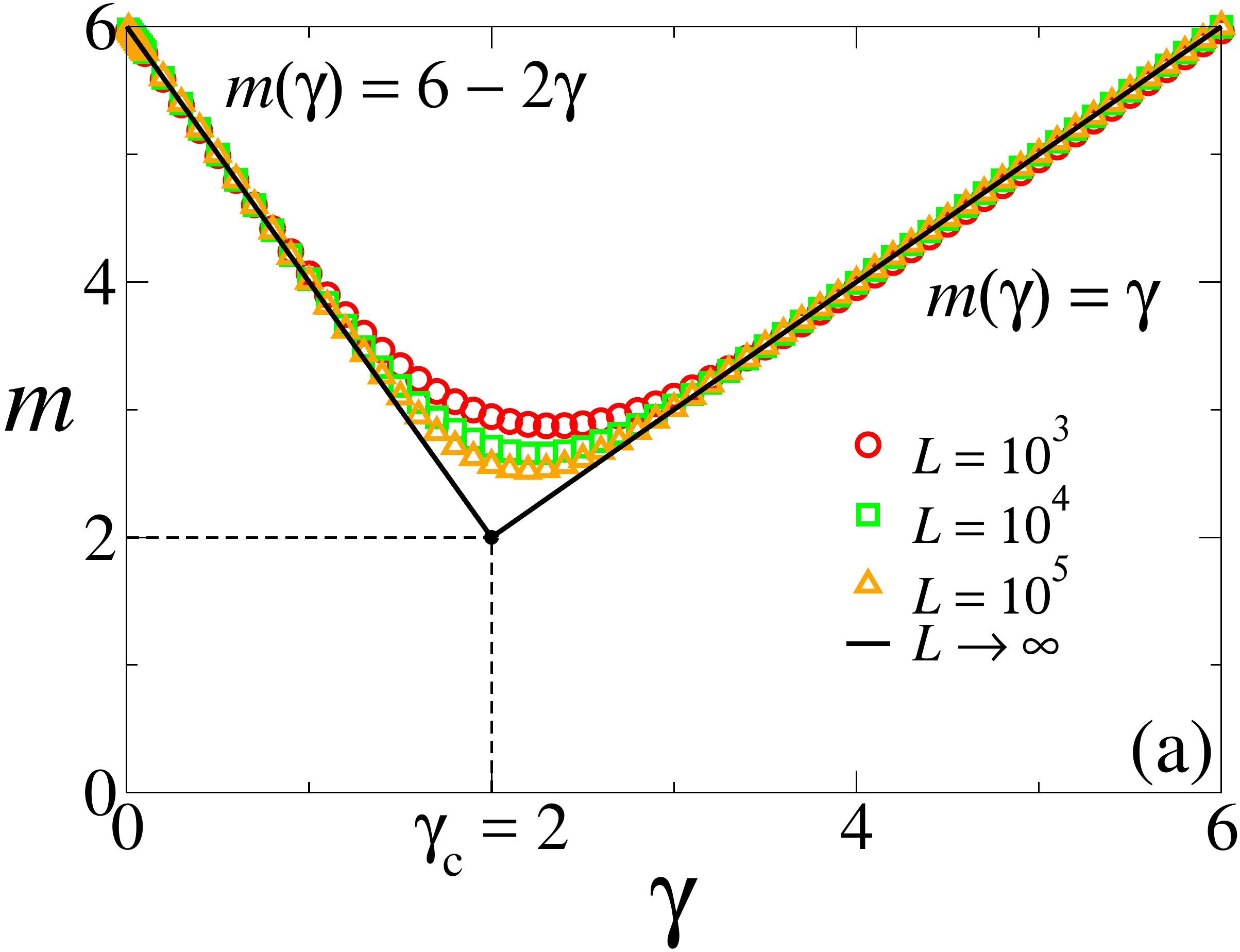} 
}\\
\subfloat{
\includegraphics[width=\columnwidth]{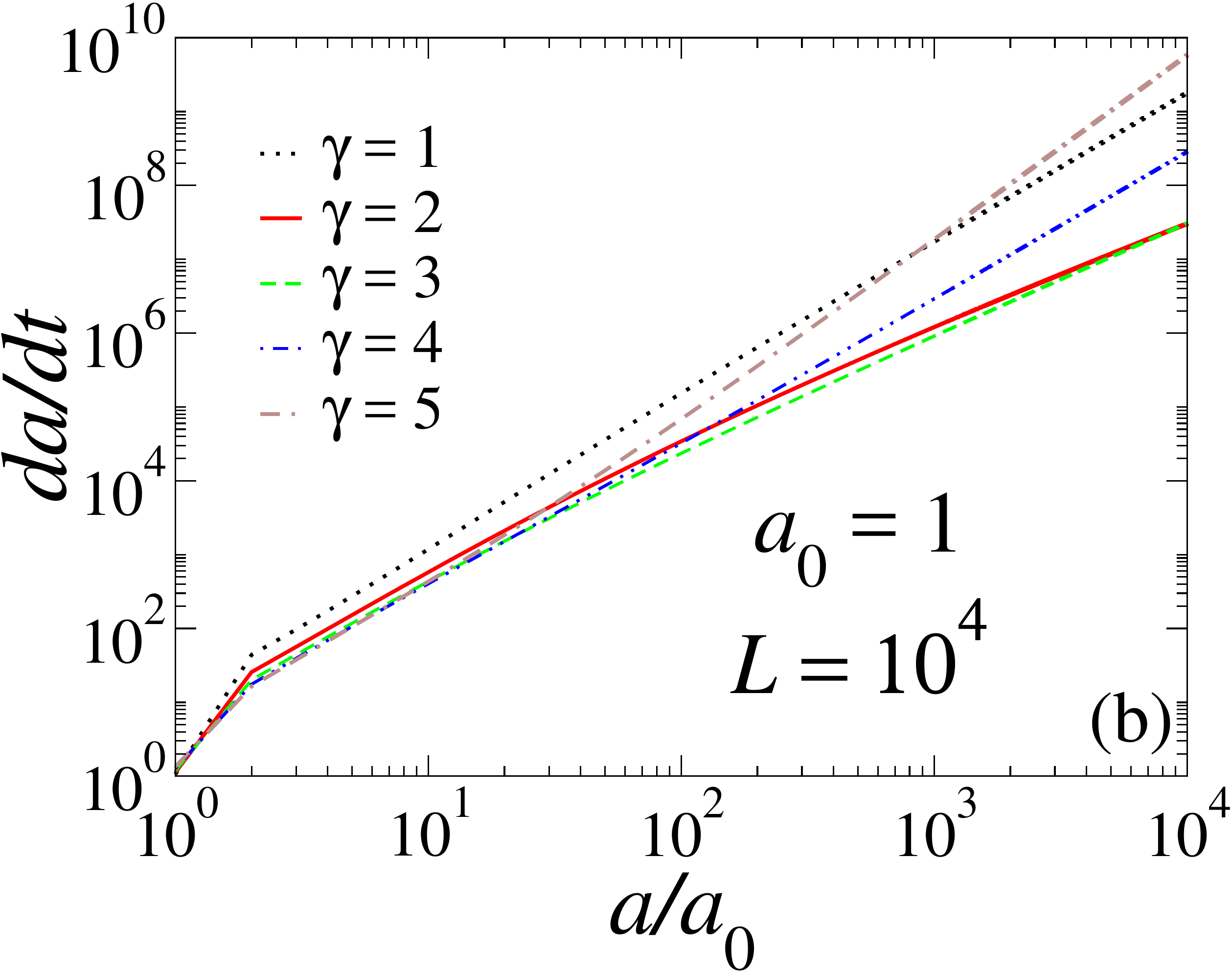} 
}
\caption{(Color online) Top: numerical dependence of the Paris exponent $m$ on the 
damage-accumulation exponent $\gamma$ for system sizes ranging from 
$L=10^3$ to $L=10^5$. The solid line corresponds to an 
extrapolation of the results to the thermodynamic limit assuming the finite size 
scaling hypothesis given by Eq. (\ref{fss}).
Bottom: typical 
curves of $da/dt$ as a function of $a/a_0$ for several values of the damage-accumulation
exponent $\gamma$.}
\label{fig4}
\end{figure}

\section{The uniform case\label{sec:uniform}}

When all fatigue thresholds are equal, i.e. in the uniform limit, 
the monotonic behavior of the stress amplitude function $\Delta\sigma(x;a)$ 
(see Fig. \ref{fig2}) ensures the existence of a single 
crack along the whole rupture process. Furthermore, the crack
always advances symmetrically, with elements at both crack tips
breaking simultaneously.
As already shown in Ref. \cite{Vieira2008}, the iteration of 
Eqs. (\ref{damage}) and (\ref{recurrence}), along with
the crack-growth condition, lead to a crack
growth dynamics reproducing the Paris law, as illustrated in 
Fig. \ref{fig4}.

\begin{figure}
\centering
\subfloat{
\includegraphics[width=\columnwidth]{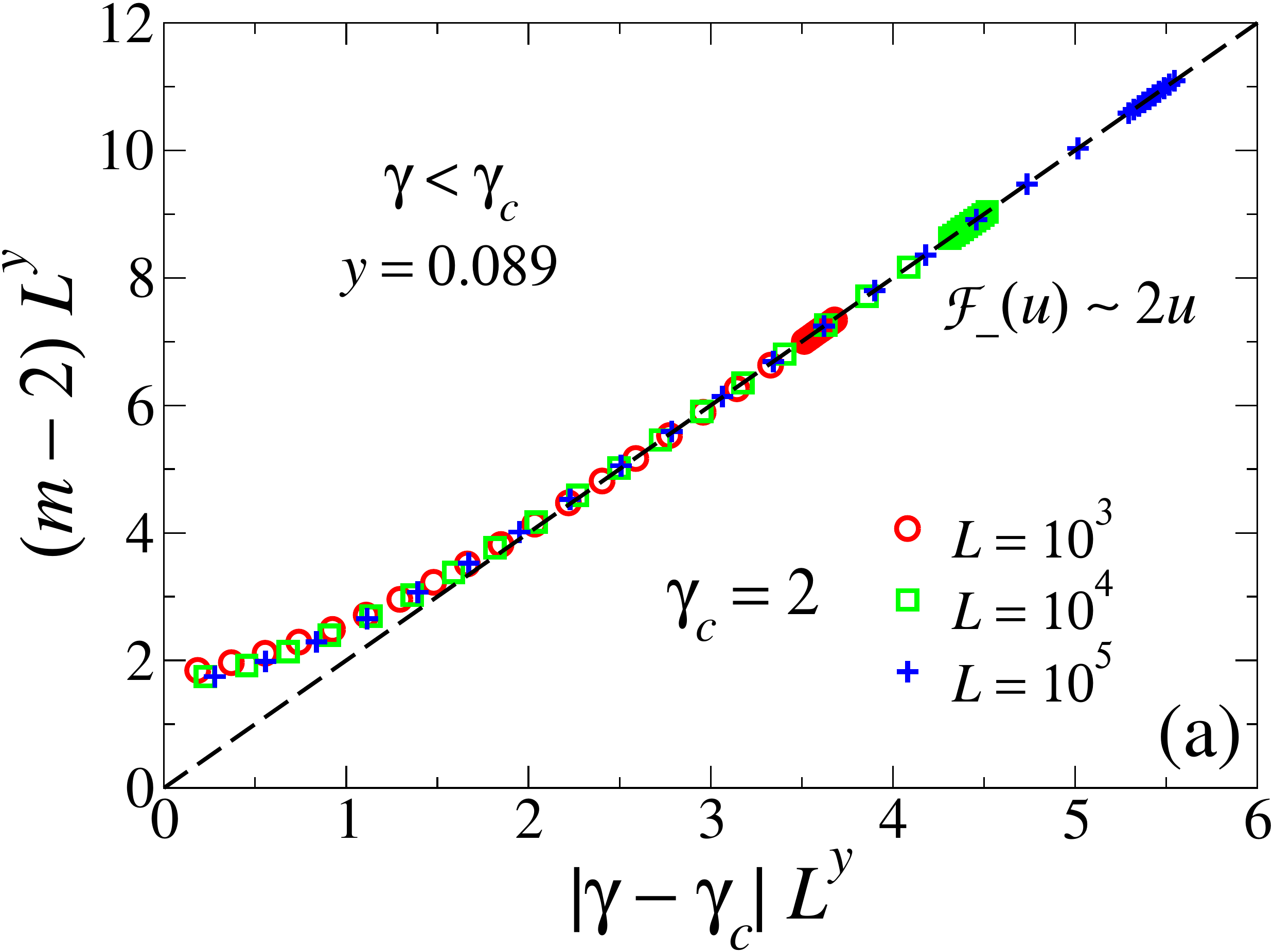} 
}\\
\subfloat{
\includegraphics[width=\columnwidth]{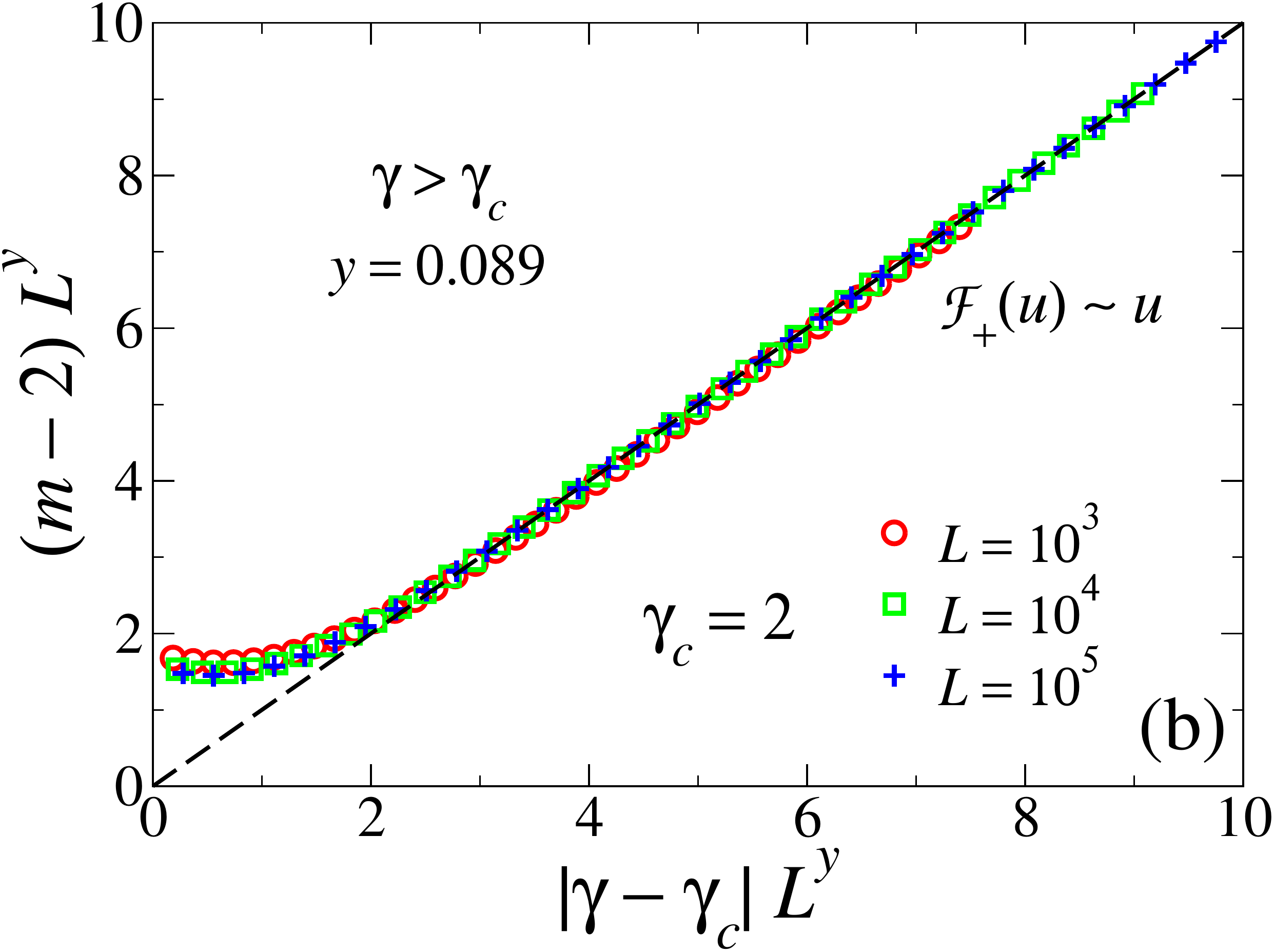} 
}
\caption{(Color online) 
Scaling plots of the dependence of $m$ on $\gamma$
and $L$, following Eq. (\ref{fss}),
for different 
system sizes ranging from $L=10^3$ to $L=10^5$. 
Top: $\gamma<\gamma_c=2$. Bottom: $\gamma>\gamma_c=2$.
}
\label{fig5}
\end{figure}

In the thermodynamic limit (i.e. for system sizes $L\rightarrow\infty$), the relation
between the Paris exponent $m$ and the damage-accumulation exponent
$\gamma$ is a piecewise-linear function
\begin{align}
 \begin{array}{cccc}
 m(\gamma)=
 \left\{
  \begin{array}{ll}
  6-2\gamma, \ \ \gamma\le\gamma_c;\\
  \gamma,\ \ \gamma>\gamma_c,\\
  \end{array}
 \right.
 \end{array}
\end{align}
with
\[
 \gamma_c = 2.
\]
This follows from both analytical calculations for $\gamma>\gamma_c$ 
(see below) and from 
a finite-size scaling analysis of numerical calculations,
according to
\begin{gather}
 \begin{array}{cccc}
 m(\gamma;L)-\gamma_c=
 \left\{
  \begin{array}{ll}
  L^{-y}\,\mathcal{F}_-(|\gamma-\gamma_c|L^y),\ \ \gamma<\gamma_c,\\
  L^{-y}\,\mathcal{F}_+(|\gamma-\gamma_c|L^y),\ \ \gamma>\gamma_c.
  \label{fss}\\
  \end{array}
 \right.
 \end{array}
\end{gather}
Here the system size $L$ is the number of discretized elements up to which
the calculations are iterated, $\mathcal{F}_{\pm}$ 
are scaling functions 
and $y$ is an exponent to be determined from the best data
collapse of properly rescaled plots
according to Eq. (\ref{fss}). 
As shown in Fig. \ref{fig5}, this finite-size scaling
hypothesis is nicely reproduced by numerical data
for all values of $\gamma$.

\begin{figure}
\centering
\subfloat{
\includegraphics[width=\columnwidth]{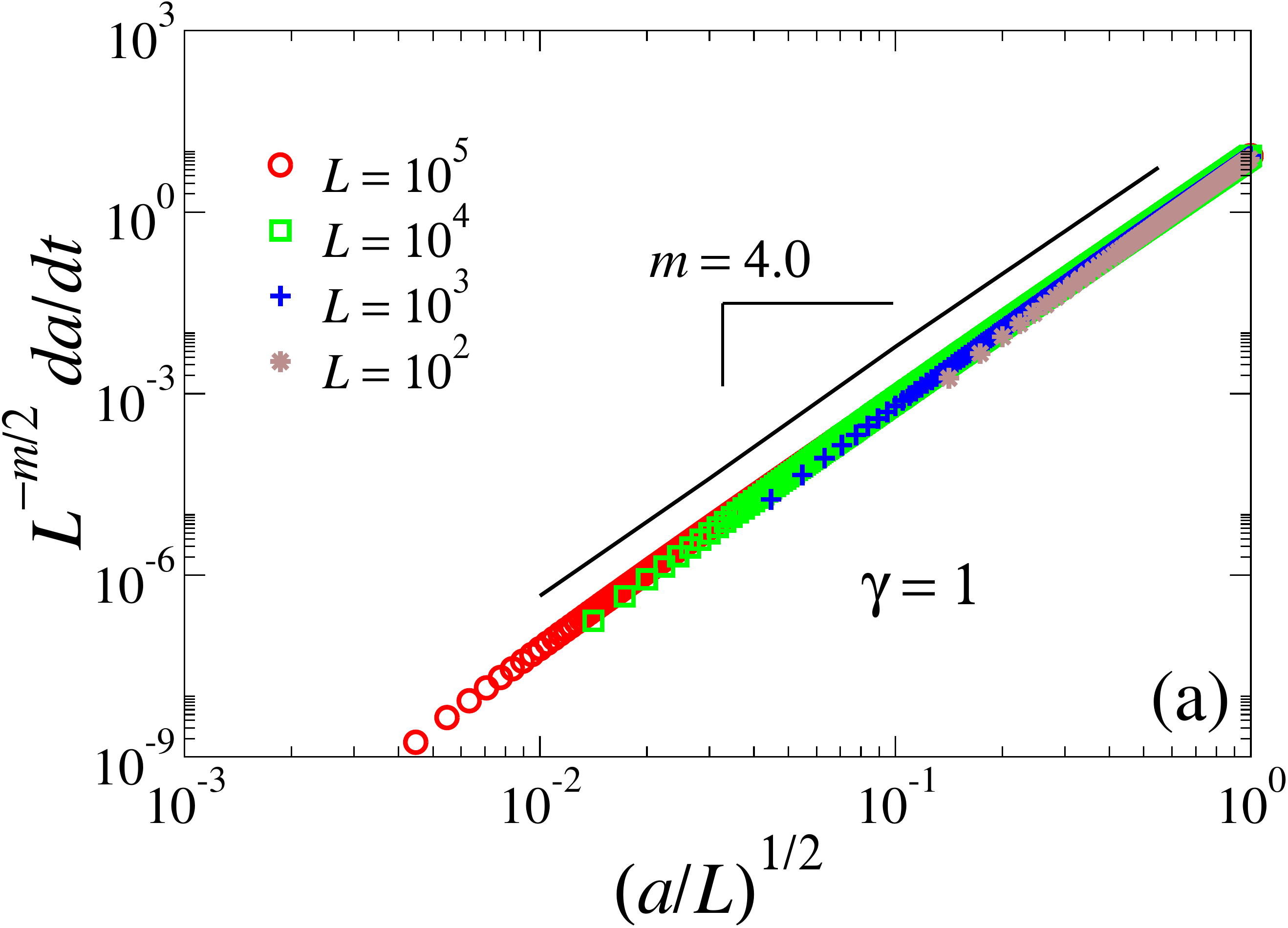} 
}\\
\subfloat{
\includegraphics[width=\columnwidth]{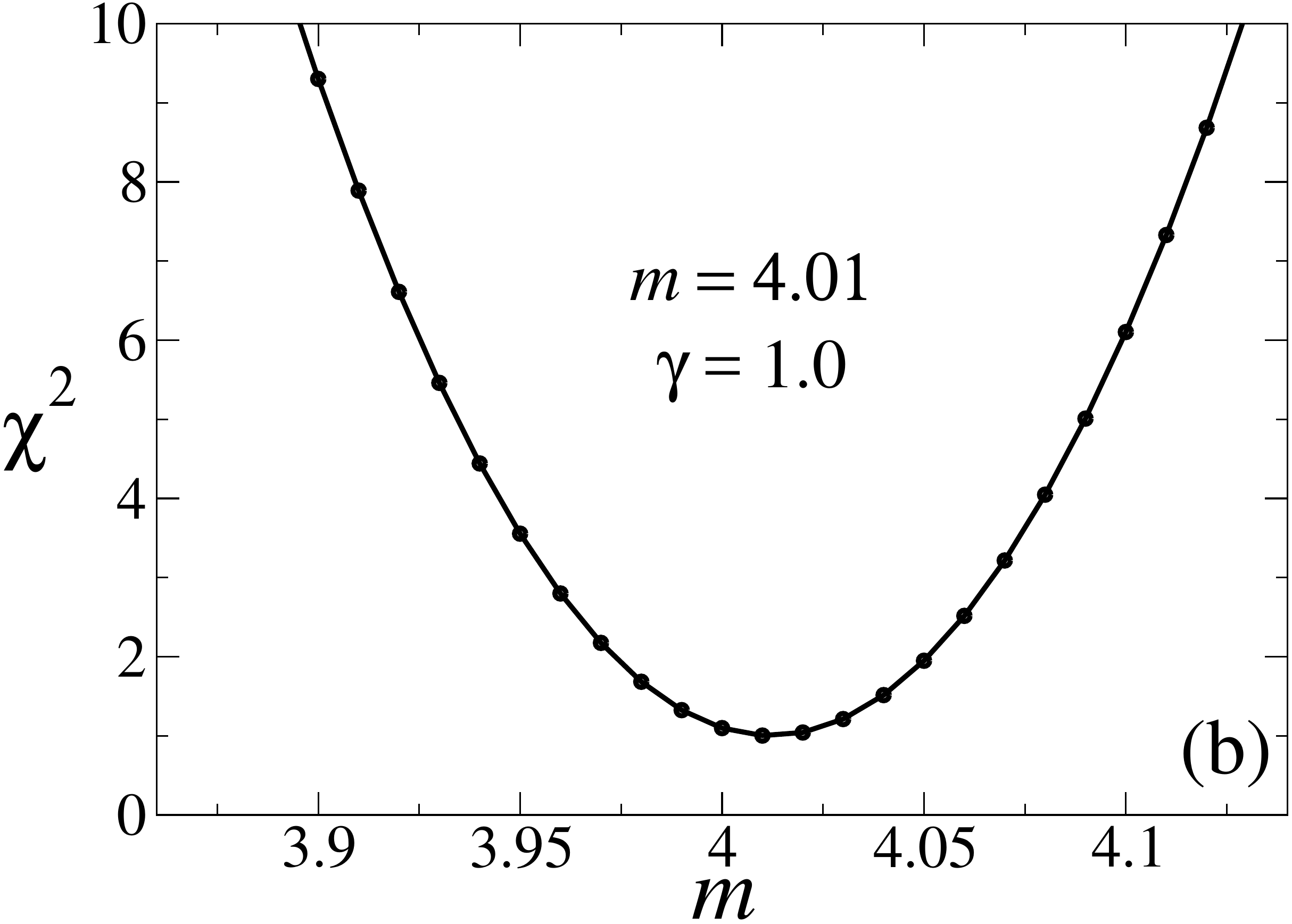} 
}
\caption{(Color online) Top: scaling plots of the dependence of
$da/dt$ on $a$ and $L$,
for different system
sizes ranging from $L=10^2$ to $L=10^5$, with $\gamma=1$.
Bottom: mean-square error of the data collapse as a 
function of the rescaling parameter $m$, showing a 
minimum very close to $m=4$. The mean-square error is calculated
by summing squares of relative deviations of rescaled ordinates,
for all rescaled values of abscissas, 
between all possible pairs of data sets. 
}
\label{fig6}
\end{figure}

The critical value $\gamma_c=2$ of the damage-accumulation exponent
is related to the divergence of the stress integral along the crack
line, as shown by the analytical calculations presented below. It
separates two regimes, one dominated by damage accumulation mostly
around the crack tip, which happens for $\gamma\gg 1$, and one
in which damage accumulations occurs more uniformly along the crack line,
as in the limiting case $\gamma\rightarrow 0$.

An alternative method to obtain the thermodynamic limit from the numerical 
results comes from an analysis of crack tip velocity versus crack length
for a single value of $\gamma$, according to the finite-size scaling 
hypothesis 
\begin{equation}
 L^{-m/2}\,\frac{da}{dt}\sim\left(\frac{a}{L}\right)^{m/2},\label{fss2}
\end{equation}
where now $m$ is chosen so as to produce the best data collapse of the rescaled
curves, as illustrated in Fig. \ref{fig6}. This yields a continuous curve
(not shown in Fig. \ref{fig4}),
which agrees quite well with the previous piecewise linear prediction,
except in the neighborhood of $\gamma_c$, where logarithmic corrections
to a simple power-law behavior are expected to be relevant. Nevertheless, this
alternative method turns out to be less susceptible to statistical
fluctuations, and will be used to evaluate the Paris exponent in the 
presence of disorder (see Sec. \ref{sec:disorder}).

\subsection*{Analytical calculations}

A few analytical results for the uniform limit can be derived from a 
recursion relation obtained by eliminating $\delta t(a)$ using Eqs. 
(\ref{damage}) and (\ref{recurrence}) in order
to compute the accumulated damage at the crack tip for each crack length.

In the uniform limit, as both crack tips always advance a single element at
a time, after $n$ iterations the crack length is $2a_n$, with
\[
 a_n = a_0 + n \delta r,
\]
Here $a_0$ represents the initial size of the crack, which we
assume to be larger than the minimum crack size associated with
$\Delta K_\text{thr}$, the threshold value of the stress-intensity factor
at which a fatigue crack propagates at a detectable rate. For smaller
sizes, crack growth proceeds at a very slow rate (see e.g. Ref. \cite{Krupp-Book},
chapter 2), below one atomic length per loading cycle, and our mesoscopic approach is inapplicable.
Therefore, we expect that our results only apply to the Paris regime of fatigue crack propagation.

If we define
\[
F_n\,{\equiv}\,F(a_{n+1},a_{n-1}),
\]
combining Eqs. (\ref{damage}) and (\ref{recurrence}) 
with the crack growth condition leads to
the time elapsed between consecutive rupture events,
\begin{align}
\delta t(a_n)=\frac{F_\mathrm{thr}(1-G_n)}{f_{\!_0}\left[\Delta\sigma(a_{n+1};a_{n})\right]^{\gamma}},
\label{eq:deltat}
\end{align}
and to the rescaled recursion relation
\begin{align}
G_{n}\equiv\frac{F_n}{F_\mathrm{thr}}=\displaystyle\sum_{k=1}^{n}g_{nk}(1-G_{k-1})\text
{,}\ \ \ n>0, \label{recurrence-gnk}
\end{align}
\noindent with $G_{0}=0$ and
\begin{align}
g_{nk}\equiv\left[\frac{\Delta\sigma(a_0+(n+1)\delta 
r;a_0+(k-1)\delta r)}{\Delta\sigma(a_0+k\delta r;a_0+(k-1)\delta 
r)}\right]^{\gamma}.
\end{align}
Notice that $g_{nk}$ is related to the ratio between the stress amplitudes at
two different times in the rupture process, 
and that the asymptotic behavior of the rescaled accumulated damage 
$G_n$ at the crack 
tip must be taken into account in order to estimate the crack growth rate
\[
 \frac{da}{dN}\sim \frac{\delta r}{\delta t(a)}\sim \frac{1}{\delta t(a)}.
\]

\begin{figure}
\centering
\subfloat{
\includegraphics[width=\columnwidth]{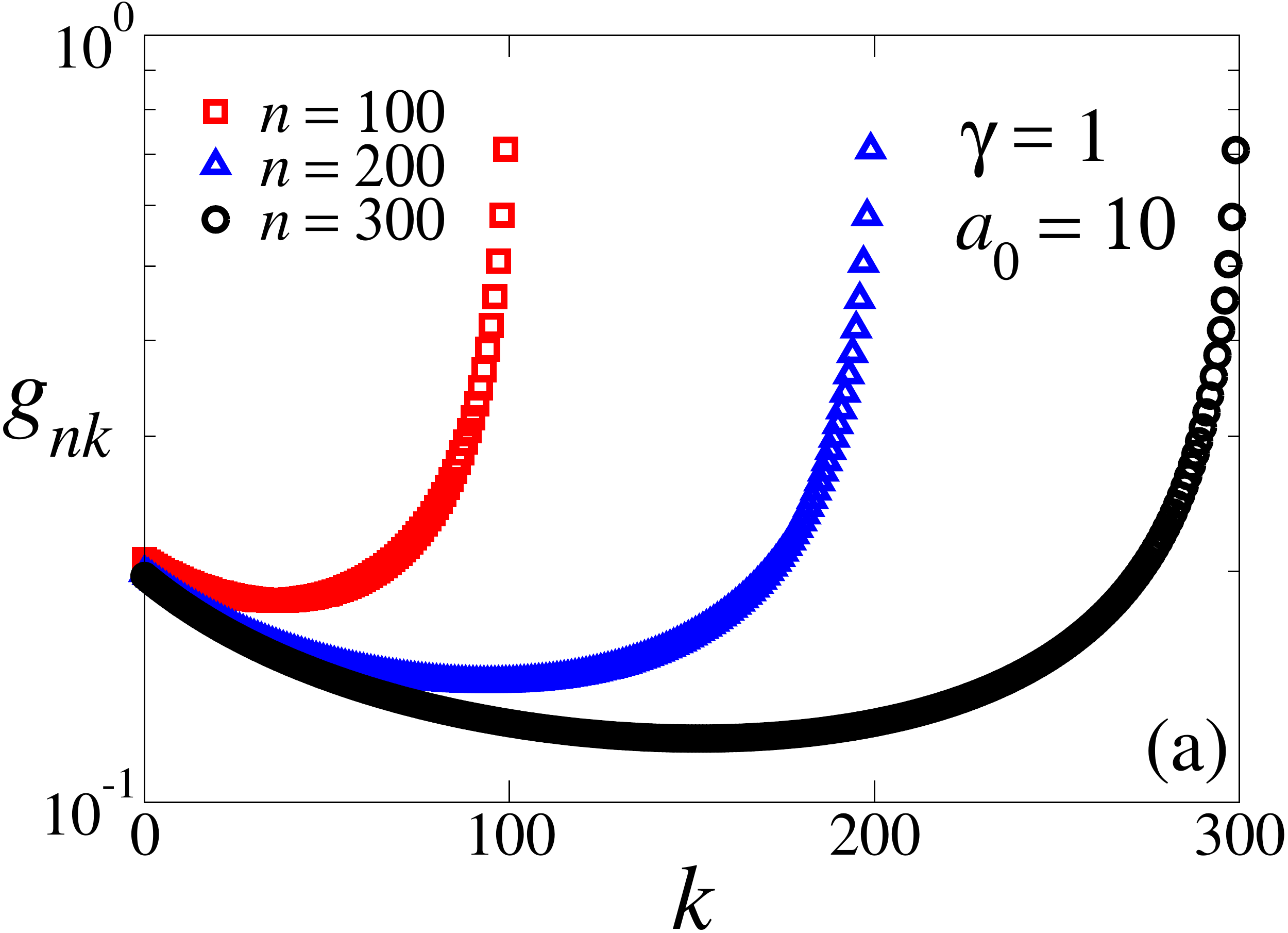} 
}\\
\subfloat{
\includegraphics[width=\columnwidth]{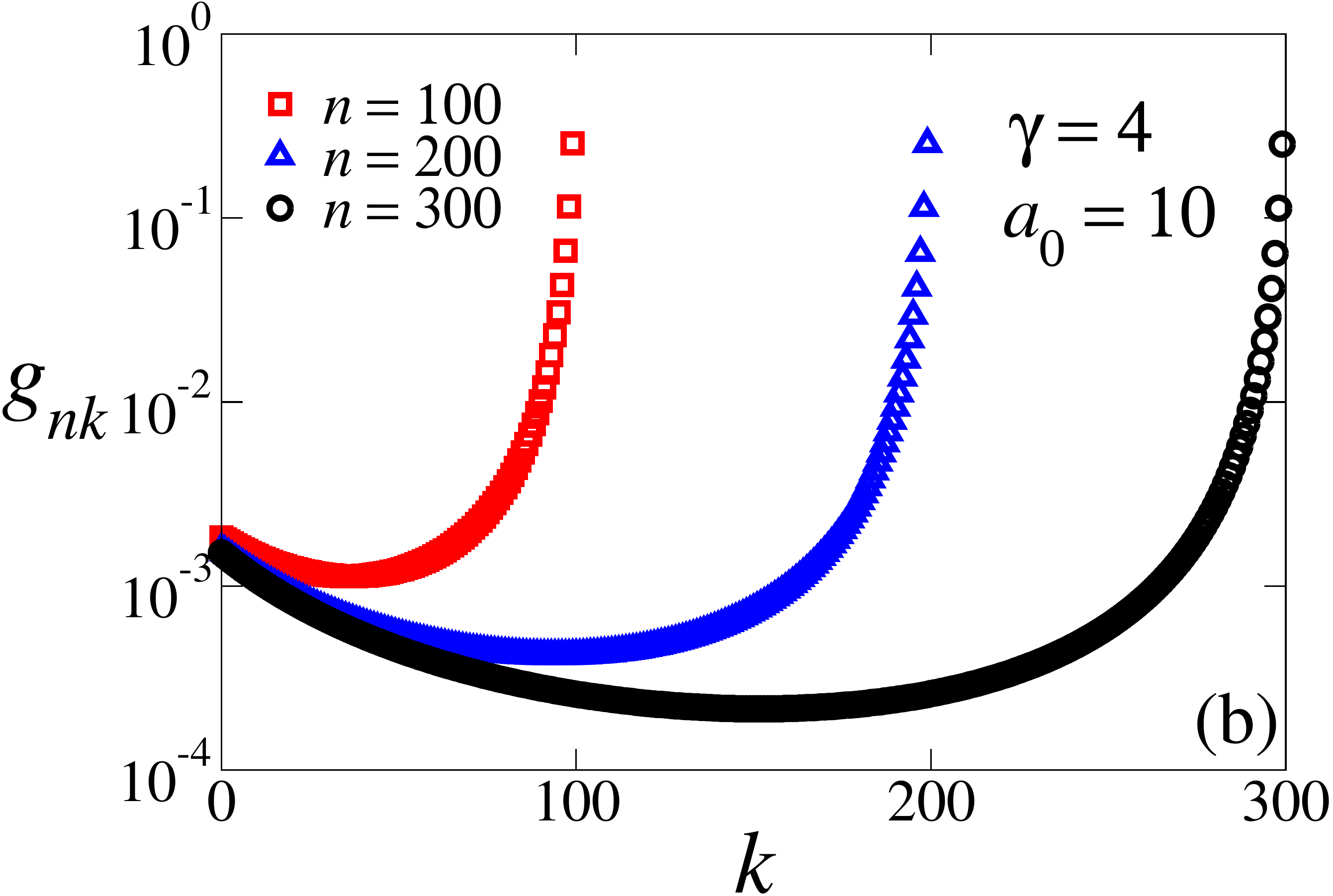} 
}
\caption{(Color online) Typical behavior of the stress amplitude amplitude ratio 
$g_{nk}$ for a few values of the damage-accumulation exponent $\gamma$. Top: 
$\gamma=1$. Bottom: $\gamma=4$. 
}
\label{fig7}
\end{figure}

The asymptotic behavior of $G_n$ is related to the asymptotic behavior
of $g_{nk}$, which is given by
\begin{align}
 \begin{array}{cccc}
 g_{nk}\sim
 \left\{
  \begin{array}{ll}
  \left(2\delta r/a_0\right)^{\gamma/2},\ \ k\delta r\ll a_0\ll n\delta r;\\
  \left(2/k\right)^{\gamma/2},\ \ a_0\ll k\delta r\ll n\delta r;\\
  \left(n-k+2\right)^{-\gamma/2},\ \ a_0\ll k\delta r\approx n\delta r. \label{gnk}
  \end{array}
 \right.
 \end{array}
\end{align}
Notice that $g_{nk}$ assumes its largest values for $k$ approaching $n$,
as shown in Fig. \ref{fig7}.

If $G_n$ approaches a value $G^*$ smaller than unity as $n\rightarrow\infty$,
it follows from Eqs. (\ref{recurrence-gnk}) and (\ref{gnk}) that
we can write 
\begin{equation}
G^{*}\approx(1-G^{*})\sum_{k=1}^{\infty}(n-k+2)^{-\gamma/2}
\equiv \left(1-G^{*}\right)s_{\infty}(\gamma),
\label{approx}
\end{equation}
with
\begin{equation}
 s_{\infty}(\gamma)=
 \left\{
  \begin{array}{ll}
  \zeta(\frac{\gamma}{2})-1,\ \ \gamma>2;\\
  \infty,\ \ \gamma\le2,\\
  \end{array}
 \right.
\end{equation}
where $\zeta(x)$ is the Riemann zeta function.

Therefore, for $\gamma>2$ we have
\begin{gather}
G^*\approx \displaystyle\frac{s_{\infty}(\gamma)}{1+s_{\infty}(\gamma)} \ \ \ 
\Rightarrow \ \ \ \frac{da_n}{dt}\approx\frac{a_n^{\gamma/2}}{1-G^*}\sim 
a_n^{\,\gamma/2}\label{g*},
\end{gather}
yielding $m=\gamma$.
However, this analysis breaks down for $\gamma<2$, since $G^*$ approaches
unity as $\gamma\rightarrow 2^+$.

Nevertheless, for $\gamma\rightarrow 0^+$ we can write 
\begin{equation}
\left[\Delta\sigma(x;a_n)\right]^\gamma=\exp\left\{\ln
\left[\Delta\sigma(x;a_n)\right]^\gamma\right\} 
\approx 1+\ln\left[\Delta\sigma(x;a_n)\right]^\gamma, 
\nonumber
\end{equation}
from which, by using Eqs. (\ref{recurrence-gnk}) and (\ref{gnk}),
we obtain, for $n>1$,
\begin{equation}
G_{n}\approx1+\ln\left(\frac{g_{n,1}}{g_{n-1,1}}\right)
\label{eq:exp1}
\end{equation}
and
\begin{equation}
\frac{g_{n,1}}{g_{n-1,1}}\approx 1-\gamma\left(\frac{a_0}{\delta 
r}\right)^2 n^{-3}.
\label{eq:exp2}
\end{equation}
Therefore, as $\gamma\rightarrow 0^+$ we have
\begin{equation}
 \delta t(a_n) \sim\frac{1-G_n}{a_n^{\gamma/2}} \sim 
 -\ln\left(\frac{g_{n,1}}{g_{n-1,1}}\right) \sim \gamma a_n^{-3}
 \label{eq:exp3}
\end{equation}
so that we obtain a Paris exponent $m=6$, in agreement with the
numerical results. Notice however that the multiplicative coefficient
in the Paris law expression, which in this limit is proportional
to $\gamma^{-1}$, diverges as $\gamma\rightarrow 0$, in agreement
with the expectation of sudden rupture when the damage threshold
is reached simultaneously at all points.

\section{
The uniform case with a modified damage-accumulation rule
\label{sec:modunif}
}

In analogy with modifications of the Paris law suggested by
crack-closure phenomena, related to factors such as plasticity, roughness
and oxidation, which imply an effective
reduction of the stress-intensity amplitude \cite{Krupp-Book,Elber1971}, 
the damage-accumulation rule
can be modified to accommodate a threshold stress amplitude needed
to induce local damage. This can be done by rewriting Eq. (\ref{damage})
in the form
\begin{equation}
\delta F(x;a)=f_0\delta t(a)[\Delta\sigma_\mathrm{eff}(x;a)]^\gamma,
\label{damage-mod}
\end{equation}
with an effective stress amplitude 
\begin{equation}
\Delta\sigma_\mathrm{eff}(x;a) = \Delta\sigma(x;a)-b\Delta\sigma_0, 
\end{equation}
the coefficient $b$ ($0\leq b\leq 1$) giving the strength,
relative to the external stress amplitude $\Delta\sigma_0$,
of the threshold stress amplitude below which no damage accumulation occurs. 
Notice that for $b=0$ we recover the case 
investigated in Sec. \ref{sec:uniform}, whereas $b=1$ leads
to no damage accumulation infinitely far from the crack tips.

A similar analysis to the one performed in Sec. \ref{sec:uniform} shows
that Eqs. (\ref{eq:deltat}) and (\ref{recurrence-gnk}) now read
\begin{equation}
\delta t(a_n)=\frac{F_\mathrm{thr}\,(1-G_{n})}
{f_0\left[\Delta\sigma_\mathrm{eff}(a_{n+1};a_{n})\right]^{\gamma}},
\end{equation}
and
\begin{equation}
G_{n}=\sum_{k=1}^{n}h_{nk}(1-G_{k-1}),
\label{recorrencia-2}
\end{equation}
with the $g_{nk}$ of Eq. (\ref{recurrence-gnk}) replaced by
\begin{equation}
h_{nk}\equiv\left[
\frac{\Delta\sigma_\mathrm{eff}(a_0+(n+1)\delta r;a_0+(k-1)\delta r)}
{\Delta\sigma_\mathrm{eff}(a_0+k\delta r;a_0+(k-1)\delta r)}\right]^{\gamma},
\end{equation}
whose asymptotic behavior is given by
\begin{equation}
 h_{nk}\approx
 \left\{
  \begin{array}{ll}
  \left[\frac{(1-b)\sqrt{2a_0\delta r}}
  {a_0-b\sqrt{2a_0\delta r}}\right]^{\gamma},\ \ k\delta r\ll a_0\ll 
n\delta r;\\
  \\
  \left[\frac{(1-b)\sqrt{2k}}{k-b\sqrt{2k}}\right]^{\gamma},\ \ 
a_0\ll k\delta r\ll n\delta r;\\
  \\
  {(n-k+2)^{-\gamma/2}},\ \ a_0\ll k\delta r\approx n\delta 
r.\label{assymptotic-2}
  \end{array}
 \right.
\end{equation}
Thus, Eq. (\ref{g*}) remains valid for $\gamma>\gamma_c$, and we still
have $m=\gamma$, with $\gamma_c=2$ irrespective of the value of $b$.

\begin{figure}
\centering
\subfloat{
\includegraphics[width=\columnwidth]{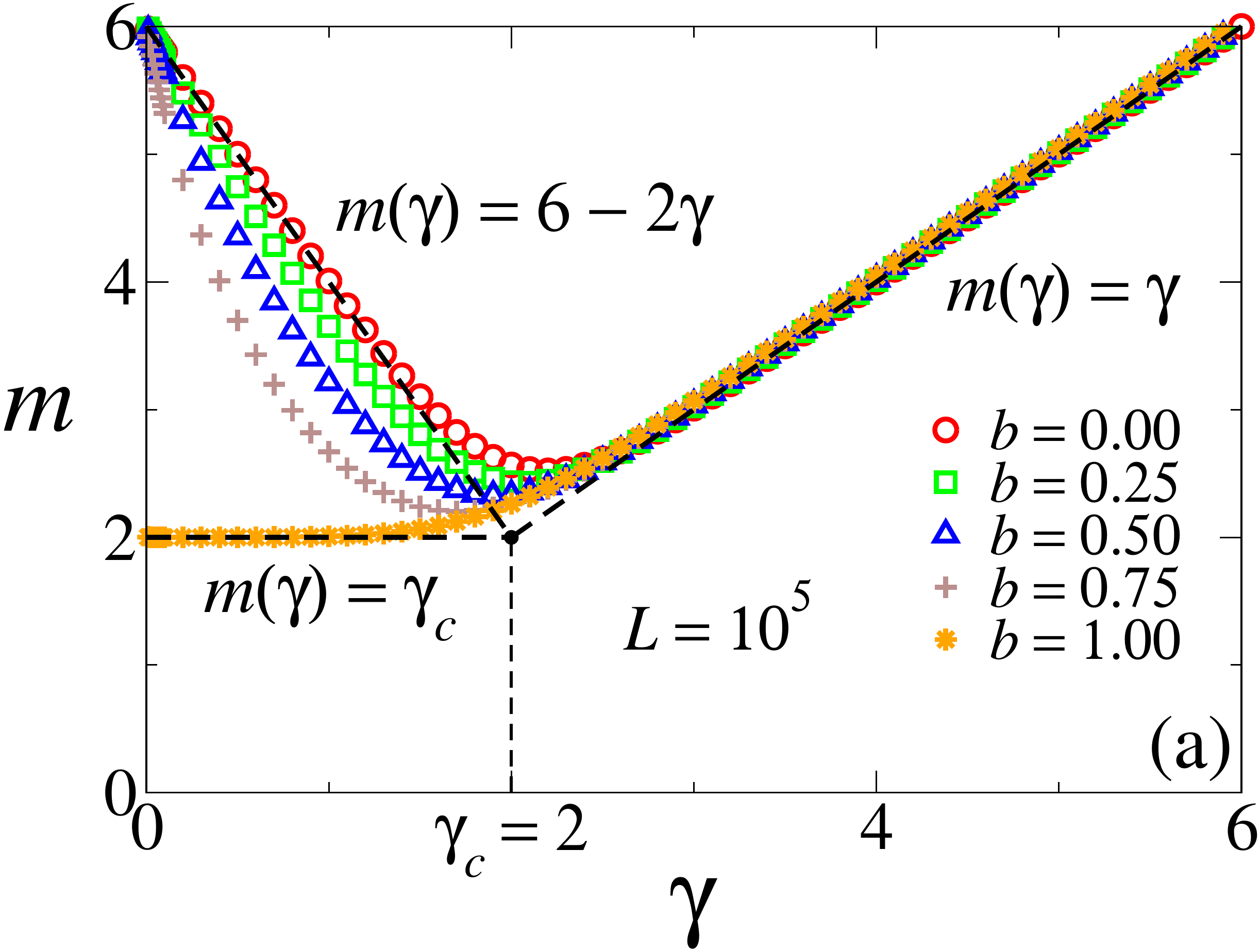} 
}\\
\subfloat{
\includegraphics[width=\columnwidth]{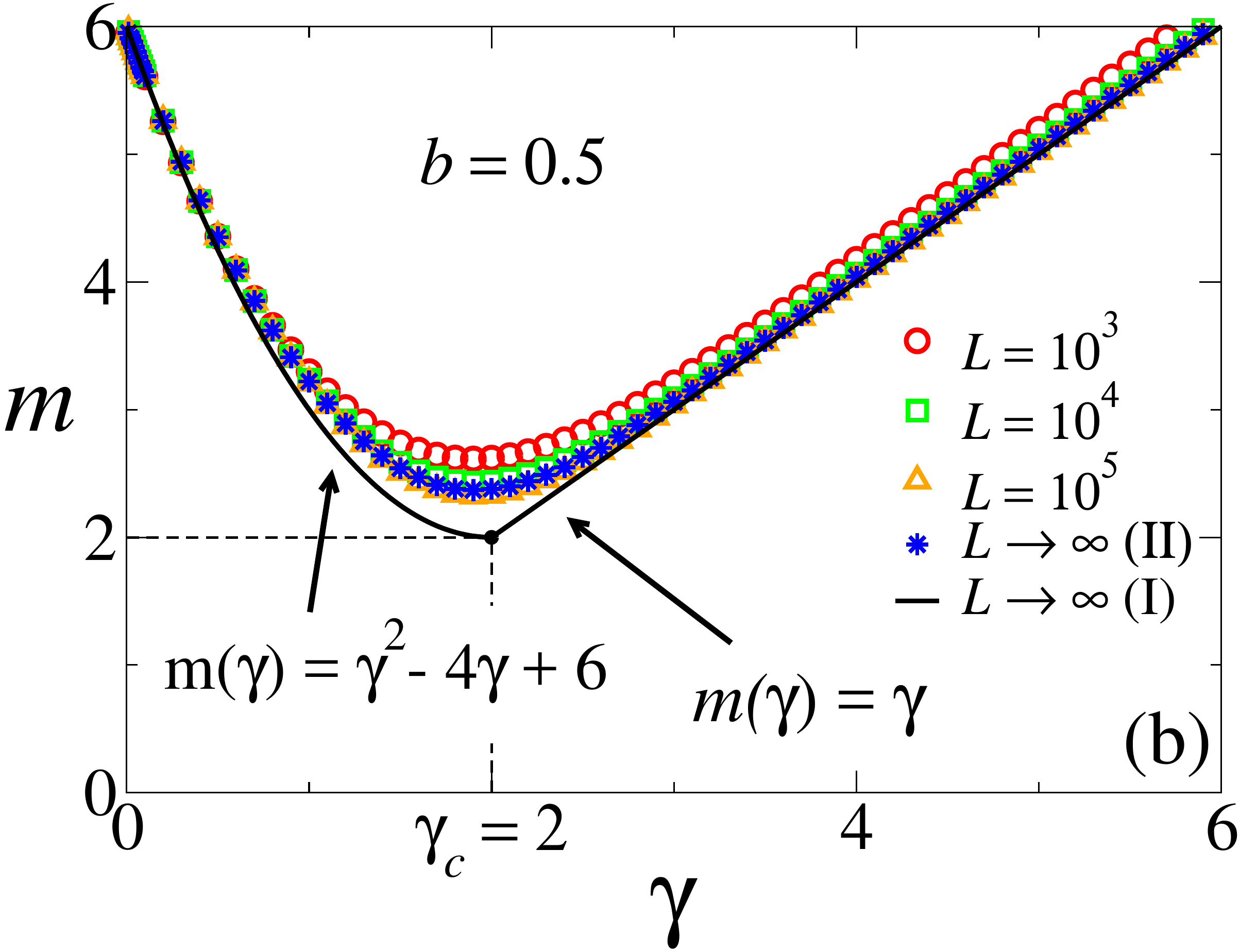}
}
\caption{(Color online) Top: numerical dependence of the Paris exponent $m$ on the 
damage-accumulation exponent $\gamma$, within the modified version of the 
model, for a few values of the threshold-stress-range parameter $b$ and system
size $L\,{=}\,10^5$. 
Notice that the linear relation $m=\gamma$ seems to be recovered for
$\gamma>2$, but a nonlinear relation seems to emerge for $\gamma<2$
if $0<b<1$. 
Bottom: finite-size behavior of $m$ against $\gamma$ for $b=0.5$. The continuous curve
is a polynomial guess for the infinite-size behavior. Blue stars indicate the 
results obtained by the alternative finite-size scaling scheme employing
Eq. (\ref{fss2}).}
\label{fig9}
\end{figure}

On the other hand, in the limit of small damage-accumulation exponent 
($\gamma\rightarrow 0^+$), the expansion in Eq. (\ref{eq:exp1})
becomes
\begin{align}
G_n\approx1+\ln\left(\frac{h_{n1}}{h_{n-1\,1}}\right), \ \ n>1.
\end{align}
Now we have to distinguish between the cases $0\leq b<1$ and $b=1$.
If $0\leq b<1$, then
\begin{align}
\frac{h_{n,1}}{h_{n-1,1}}\approx1-\gamma\left(\frac{a_0}{\delta r}
\right)^{2}\frac{1}{(1-b)}n^{-3},
\end{align}
so that
\begin{equation}
  1-G_{n}\sim \gamma a_n^{\,-3},
\end{equation}
whereas if $b=1$ we have
\begin{align}
\frac{h_{n,1}}{h_{n-1,1}}\approx1-2\gamma a_n^{-1},
\end{align}
and thus
\begin{equation}
1-G_{n}\sim \gamma a_n^{-1}.
\end{equation}
Therefore,
\begin{gather}
 m(\gamma\to0)=
 \left\{
  \begin{array}{ll}
  6, \ \ 0\leq b<1,\\
  2,\ \ b=1.
  \end{array}
 \right.
\end{gather}
Numerical calculations suggest that
for $0<\gamma<2$ a Paris regime still exists, but with a nonlinear 
relation between $m$ and $\gamma$ if $0<b<1$; see Fig. \ref{fig9}.

\section{
Introducing disorder in the fatigue thresholds
\label{sec:disorder}}

\begin{figure}
\centering
\includegraphics[width=\columnwidth]{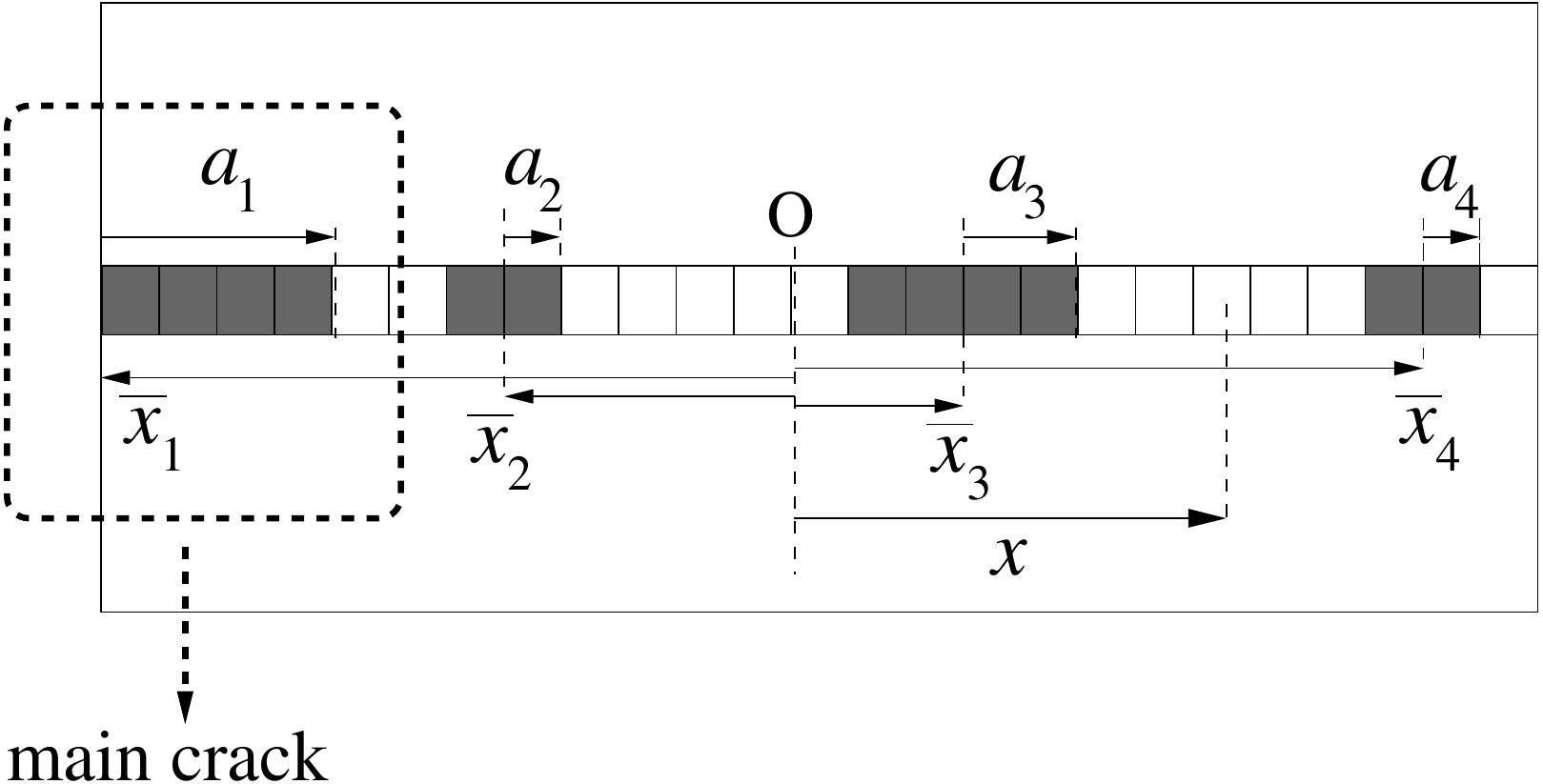}
\caption{Schematic diagram representing a configuration of the system with
random fatigue thresholds. In this case 
we observe the presence of multiple cracks (each one indicated by a sequence 
of dark elements) along the propagation line. 
}
\label{fig11}
\end{figure}

In this section we turn our attention to the description of crack growth in a 
heterogeneous medium by introducing disorder in the fatigue
thresholds. We assume that the element at position $x$ along the crack line
has a fatigue threshold $F_\mathrm{thr}(x)$ chosen randomly from
the uniform probability distribution
\begin{align}
 P(F_\mathrm{thr})=\displaystyle\frac{1}{\Delta 
F}\,\theta(F_2-F_\mathrm{thr})\theta(F_\mathrm{thr}-F_1),
\end{align}
where $\theta(x)$ is the Heaviside step function and $\Delta F\equiv 
F_2-F_1$ gauges the disorder strength, with the additional condition
that, in appropriate units, $F_1+F_2=2$. We also assume that the fatigue
thresholds at different elements are uncorrelated.

In the presence of disorder, elements far from the crack tips may reach
their fatigue thresholds, giving rise to secondary cracks,
as illustrated in Fig. \ref{fig11}. In such case,
we focus on the growth of the initial or main crack, noting that
it may coalesce with secondary cracks as the growth dynamics proceeds.

After the rupture of $n$ elements, we label the configuration of the system as
\begin{align}
 \{a_k,\overline{x}_k\}_n,
\end{align}
where $a_k$ is the half-length of the $k$th crack, which is centered at position 
$\overline{x}_k$ with respect to the midpoint of the initial crack.
We assume that between rupture events an element at position $x$ is subject to
damage accumulation following
\begin{equation}
\delta F(x;\{a_k,\overline{x}_k\}_n)=f_0\delta 
t(\{a_k,\overline{x}_k\}_n)[\Delta\sigma(x;\{a_k,\overline{x}_k\}_n)]^{\gamma},
\end{equation}
where $\delta t(\{a_k,\overline{x}_k\}_n)$ is the time elapsed between
the $n$th and the $(n+1)$th rupture events, 
and $\Delta\sigma(x;\{a_k,\overline{x}_k\}_n)$ is the 
corresponding stress amplitude at position $x$. This is analogous to 
Eq. (\ref{damage}), so that, in the notation of Sec. \ref{sec:modunif}, we
take $b=0$.

As a rupture event involves the element requiring the least
time to reach its fatigue threshold, the analogue of 
Eq. (\ref{recurrence}) allows us to write 
$\delta t(\{a_k,\overline{x}_k\}_n)$ as
\begin{align}
 \delta t(\{a_k,\overline{x}_k\}_n)=\min_{x}\left\{
\frac{F_\mathrm{thr}(x)-F(x;\{a_k,\overline{x}_k\}_{n-1})}
{f_0[\Delta\sigma(x;\{a_k,\overline{x}_k\}_n)]^{\gamma}}
\right\}.\label{time-disorder}
\end{align}

It should be emphasized that, as soon as the first secondary crack appears,
the stress amplitude $\Delta\sigma(x;\{a_k,\overline{x}_k\}_n)$ is no 
longer given by the analogue of the simple form in Eq. 
(\ref{stress-range}). Due to the lack of an analytical solution for
the stress field of multiple thin cracks, even in the simplest
case where the cracks are arranged along the same line, we
resort to an independent-crack approximation, to be detailed below,
whenever it is necessary to deal with secondary cracks, except
in the case $\gamma=0$, which we now present in detail.

\subsection{The case $\gamma=0$}

In this limit, damage accumulation is independent of the local stress amplitude,
so that the problem is similar to a 1D percolation process,
and it is possible to obtain analytical results. 
In this subsection only, in order to simplify the calculations, we
assume that the initial crack is a notch of length $a_0$ at the left end 
of the medium. The case of a central initial crack was briefly discussed
in Ref. \cite{Oliveira2012}.

The probability 
of finding the main crack with length $a$ at time $t$ is given by
\begin{align}
P\left(a\left|a_0,t\right.\right)=\left[p(t)\right]^{a-a_0}\left[1-p(t)\right],
\end{align}
in which $p(t)$ is
the probability that an element has reached its fatigue threshold before
time $t$, the factor $1-p(t)$ being the probability
that the element at the (right) tip of the main crack remains intact
at time $t$.
Since for $\gamma=0$ we have $F(x;\{a_k,\overline{x}_k\}_n)=f_0 t$,
it follows that 
\begin{align}
p(t)=\min\left\{1,\frac{t-t_1}{T}\theta
\left(t-t_1\right)\right\},\label{probability}
\end{align}
where $t_1=F_1/f_0$ and $T=\Delta F/f_0$ are parameters
related to the disorder distribution.

For a semi-infinite medium, the average length of the main crack 
at time $t$ is given by
\begin{align}
\langle a\rangle_t=\sum_{a=a_0}^{\infty}aP\left(a\left|a_0,t\right.\right)=
a_0+\frac{p(t)}{1-p(t)}, 
\label{eq:at}
\end{align}
so that, eliminating $t$ from Eqs. (\ref{probability}) and (\ref{eq:at}),
the average tip velocity of the main crack can be written,
for $t_1<t<t_2\equiv F_2/f_0$, as
\begin{align}
\langle v\rangle_t=\frac{d}{dt}\langle 
a\rangle_t\sim \langle a\rangle_t^2, \label{last}
\end{align}
implying a Paris exponent $m=4$ instead of $m=6$ as in the uniform limit.

\begin{figure}[b]
\centering
\subfloat{
\includegraphics[width=\columnwidth]{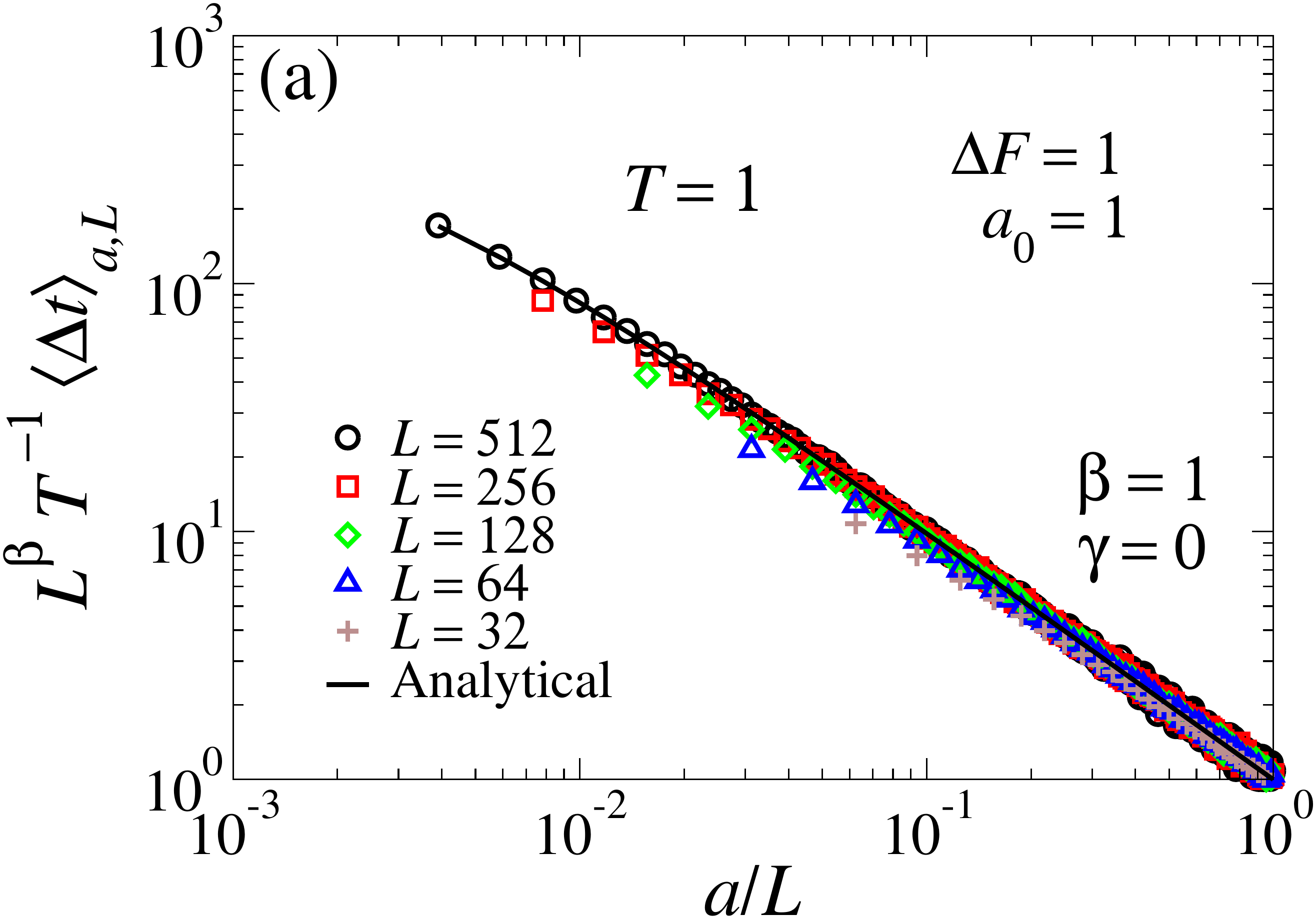} 
}\\
\subfloat{
\includegraphics[width=\columnwidth]{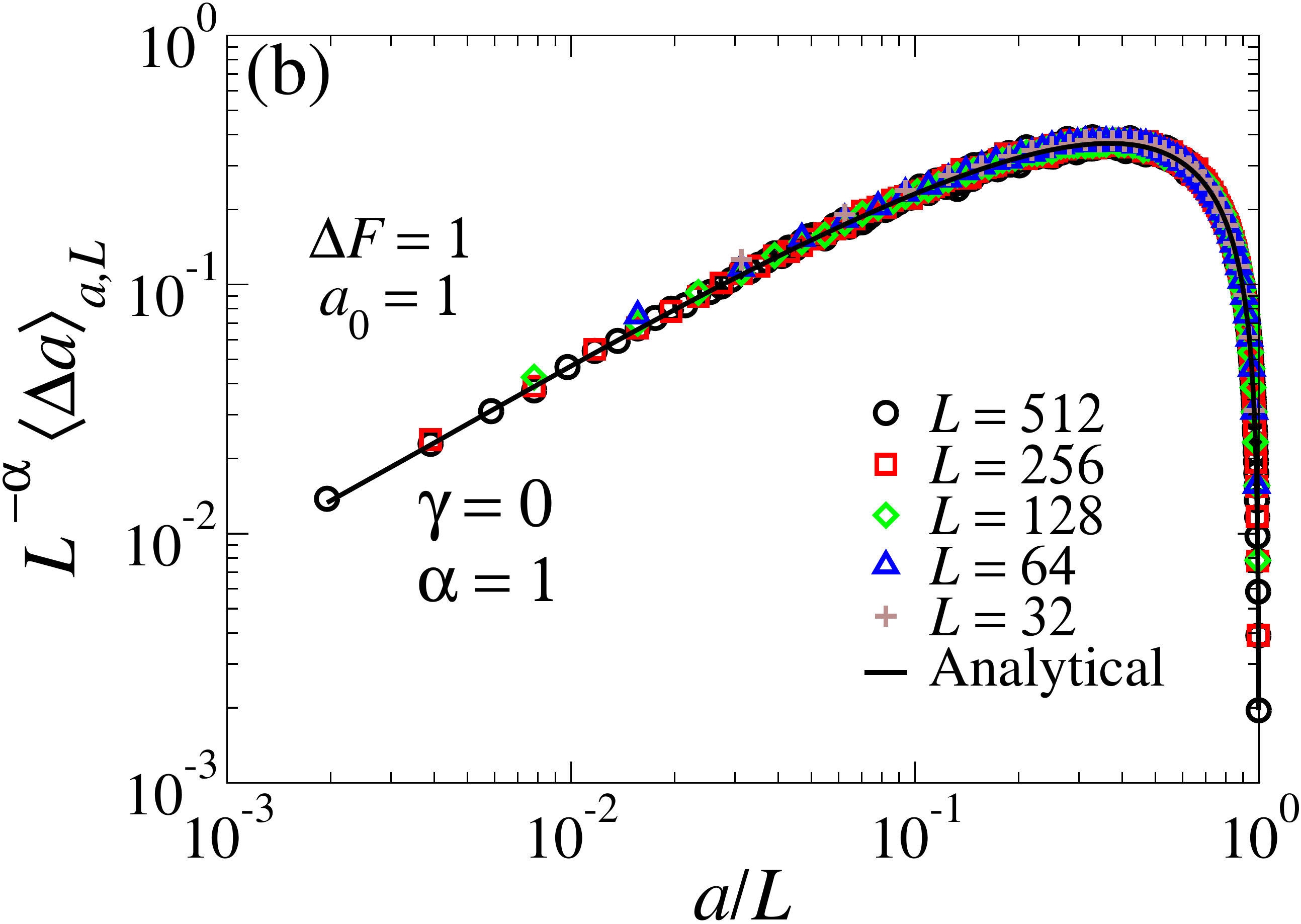}
}
\caption{(Color online) Rescaled mean values of waiting times (top) and 
avalanche sizes
(bottom) between consecutive jumps of the main crack for the disordered version of 
the model with $\gamma\,{=}\,0$. Numerical 
results are in good agreement with the analytical results from Eqs. 
(\ref{mean-da}) and (\ref{mean-dt}), indicating power-law behaviors of both
quantities as functions of the length of the main crack, in the limit of 
infinite system size.}
\label{fig12}
\end{figure}

It is also possible to study finite systems containing $L$ elements, and have access
to the distribution of waiting times between rupture events, as well as 
to the distribution of
avalanche sizes. An avalanche is defined as a sudden event in which the crack tip 
advances by more than a single discretized elements, while the avalanche
size is the number of elements by which the main crack grows
in a single event 
\footnote{Notice that an avalanche involves stress rearrangements, by changing the 
configuration of the cracks in the system. In the limit of $\gamma=0$ this stress rearrangement
is irrelevant for damage accumulation, and avalanches are just random nucleations. This is not
the case for any $\gamma>0$, and avalanche events will be correlated.}. 
To this end, we must consider the probability that the main
crack has length $a$ and, upon rupture of the element at its tip, 
happening between times $t$ and $t+dt$, advances
$\Delta a$ elements having waited a time between $\Delta t$ 
and $\Delta t+d(\Delta t)$ since it last advanced.
Denoting this probability by $\rho_{_L}\left(\Delta a,\Delta 
t,t\left|a\right.\right)\,dt\,d(\Delta t)$, we have
\begin{eqnarray}
&\rho_L\left(\Delta a,\Delta t,t\left|a\right.\right)=
\frac{(a-a_0)(a-a_0+1)}{T^2}\left[p(t-\Delta t)\right]^{a-a_0-1}
\nonumber\\&\times\left[p(t)\right]^{\Delta a-1}
\left\{\left[1-p(t)\right]\left(1-\delta_{\Delta a, L-a}\right)+
\delta_{\Delta a, L-a}\right\},
\label{eq:rhoL}
\end{eqnarray}
$\delta_{i,j}$ being the Kronecker delta symbol. Here, 
$\left[p(t-\Delta t)\right]^{a-a_0-1}$ is the probability that
$a-a_0-1$ elements are broken at time $t-\Delta t$, 
$d(\Delta t)/T$ is the probability that the previous growth
event of the main crack has occurred between times $t-\Delta t$ and 
$t-\Delta t+d\left(\Delta t\right)$, 
$dt/T$ is the probability that the new growth
event of the main crack occurs between times $t$ and $t+dt$, and
$[p(t)]^{\Delta a-1}$ is the probability that the first
$\Delta a-1$ elements to the right of the element at the crack tip
are broken before time $t$. The terms between curly brackets in
Eq. (\ref{eq:rhoL}) distinguish the case in which the crack 
stops before reaching the right end of the medium, which occurs
with probability $1-p(t)$, from the case in which catastrophic
failure occurs, corresponding to $\Delta a=L-a$. The prefactor
on the right-hand side of Eq. (\ref{eq:rhoL}) ensures
normalization.

The marginal probabilities for avalanche sizes and waiting
times are obtained from $\rho_L\left(\Delta a,\Delta t,t\left|a\right.\right)$
by integrating over the appropriate variables. 
The marginal probability for avalanche sizes $\Delta a$ is given by
\begin{widetext}
\begin{align}
P_L\left(\Delta a\left|a\right.\right) = 
\int_{t_1}^{t_2}dt\int_0^{t-t_1}d(\Delta t)
\rho_L\left(\Delta a,\Delta t,t\left|a\right.\right)
= (a{-}a_0+1)\left[\frac{1-\delta_{\Delta a, L-a}}{(\Delta 
a+a-a_0+1)(\Delta a+a-a_0)}+\frac{\delta_{\Delta a,L-a}}{L-a_0}\right],
\end{align}
while the marginal probability for waiting times between consecutive jumps is
\begin{equation}
P_L\left(\Delta t\left|a\right.\right)=
\sum_{\Delta a=1}^{L-a}\int_{t_1}^{t_1+\Delta t}dt\rho_L
\left(\Delta a,\Delta t,t\left|a\right.\right)
=\frac{(a-a_0+1)}{T}\left(1-\frac{\Delta t}{T}\right)^{a-a_0}.
\end{equation}
The mean values of avalanche sizes, $\langle\Delta a\rangle_{a,L}$, and waiting 
times, $\langle\Delta t\rangle_{a,L}$, can be computed from the above marginal 
probabilities, yielding
\begin{equation}
\langle\Delta 
a\rangle_{a,L}=(a-a_0+1)\left[\left(H_{L-a_0}-H_{a-a_0}\right)\left(1-\delta_{
a,a_0}\right)+H_{L-a_0}\delta_{a,a_0}\right],\label{mean-da}
\end{equation}
\end{widetext}
where $H_n$ is the harmonic number of order $n$, and
\begin{eqnarray}
 \langle\Delta t\rangle_{a,L}=\frac{T}{a-a_0+2}.\label{mean-dt}
\end{eqnarray}
Figure \ref{fig12} compares these last results with numerical simulations
implementing the crack growth dynamics in the limit $\gamma=0$.

\begin{figure}[b]
\centering
\includegraphics[width=\columnwidth]{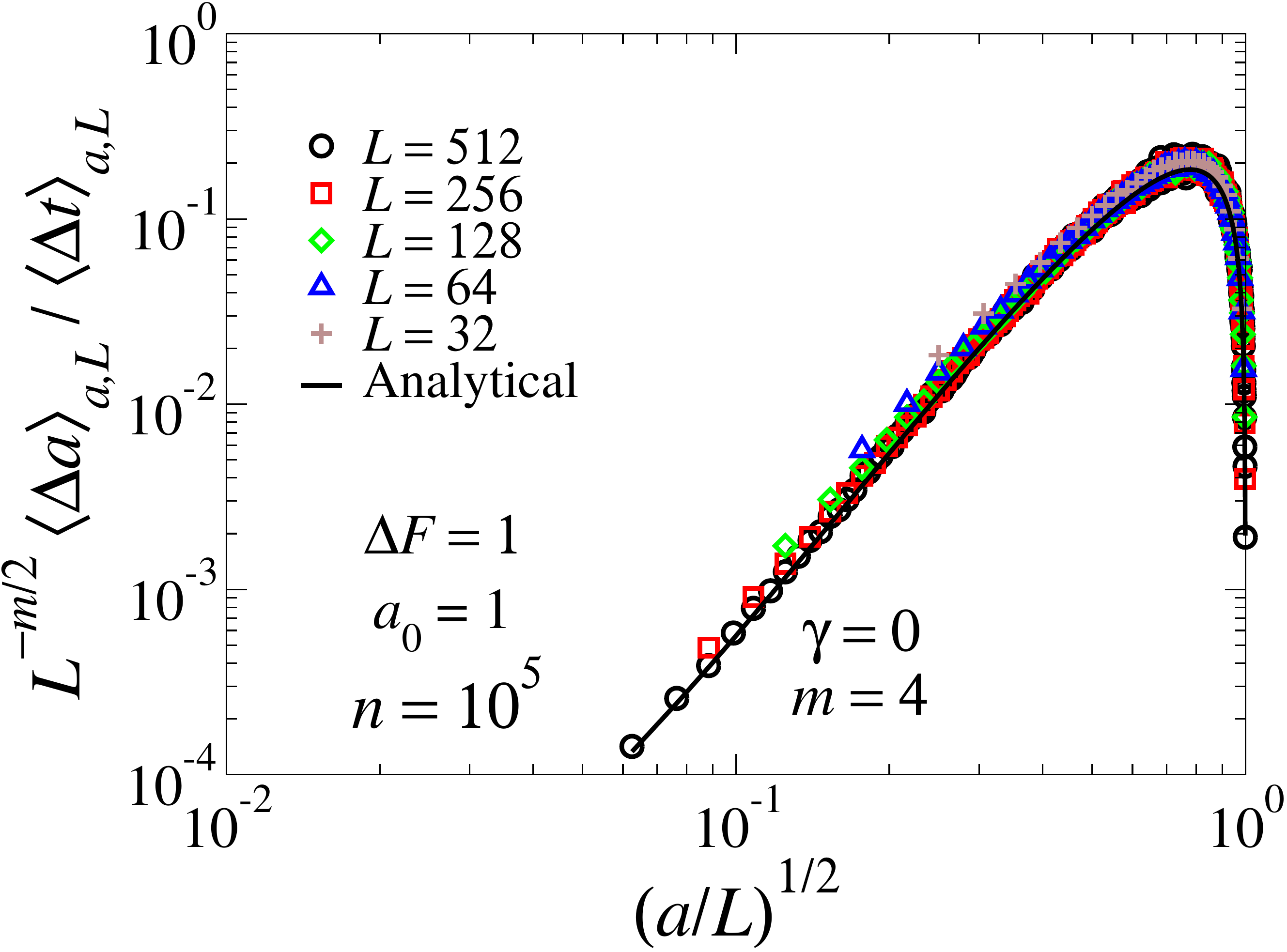}
\caption{(Color online) Rescaled mean crack-growth rate defined as the ratio 
between the mean values of avalanche size and waiting time between consecutive 
jumps for the disordered version of the model with $\gamma\,{=}\,0$. 
Numerical results are in good 
agreement with the analytical prediction of Eq. (\ref{velocity-2}),
indicating a Paris exponent equal to $m\,{=}\,4$
in the limit of infinite system size.}
\label{fig13}
\end{figure}

The ratio between those mean values yields an estimate of the crack-growth
rate, proportional to the 
the crack tip velocity of the main crack, which we define as 
\begin{align}
\langle v\rangle_{a,L}&=\frac{\langle \Delta 
a\rangle_{a,L}}{\langle \Delta 
t\rangle_{a,L}}=\frac{(a-a_0+2)(a-a_0+1)}{T}\times\nonumber\\
&\times\left[(1-\delta_{a,a_0})(H_{L-a_0}{-}H_{a-a_0})+\delta_{a,a_0}
H_{L-a_0}\right].\label{velocity-2}
\end{align}
Thus, in the limit of large crack lengths ($L\gg a \gg a_0$), we 
obtain
\begin{align}
&\langle v\rangle_{a,L}\sim a^2\ln\left(\frac{L}{a}\right),
\end{align}
leading to a Paris law with exponent $m=4$, apart from 
logarithmic corrections depending on the system size. 
Numerical simulations of the model are in 
good agreement with the analytical calculations, 
as shown in Fig. \ref{fig13}.

\subsection{The case $\gamma>0$}

In this subsection we study the properties of the disordered
model in situations 
where the damage-accumulation exponent is nonzero, a case in which
a fully analytical treatment is impossible. The 
approach we employ is therefore mostly numerical, and based on an 
\emph{independent-crack approximation} which 
neglects the correlations between the multiple cracks emerging along the 
propagation line during the breaking process.

\begin{figure}
\centering
\subfloat{
\includegraphics[width=\columnwidth]{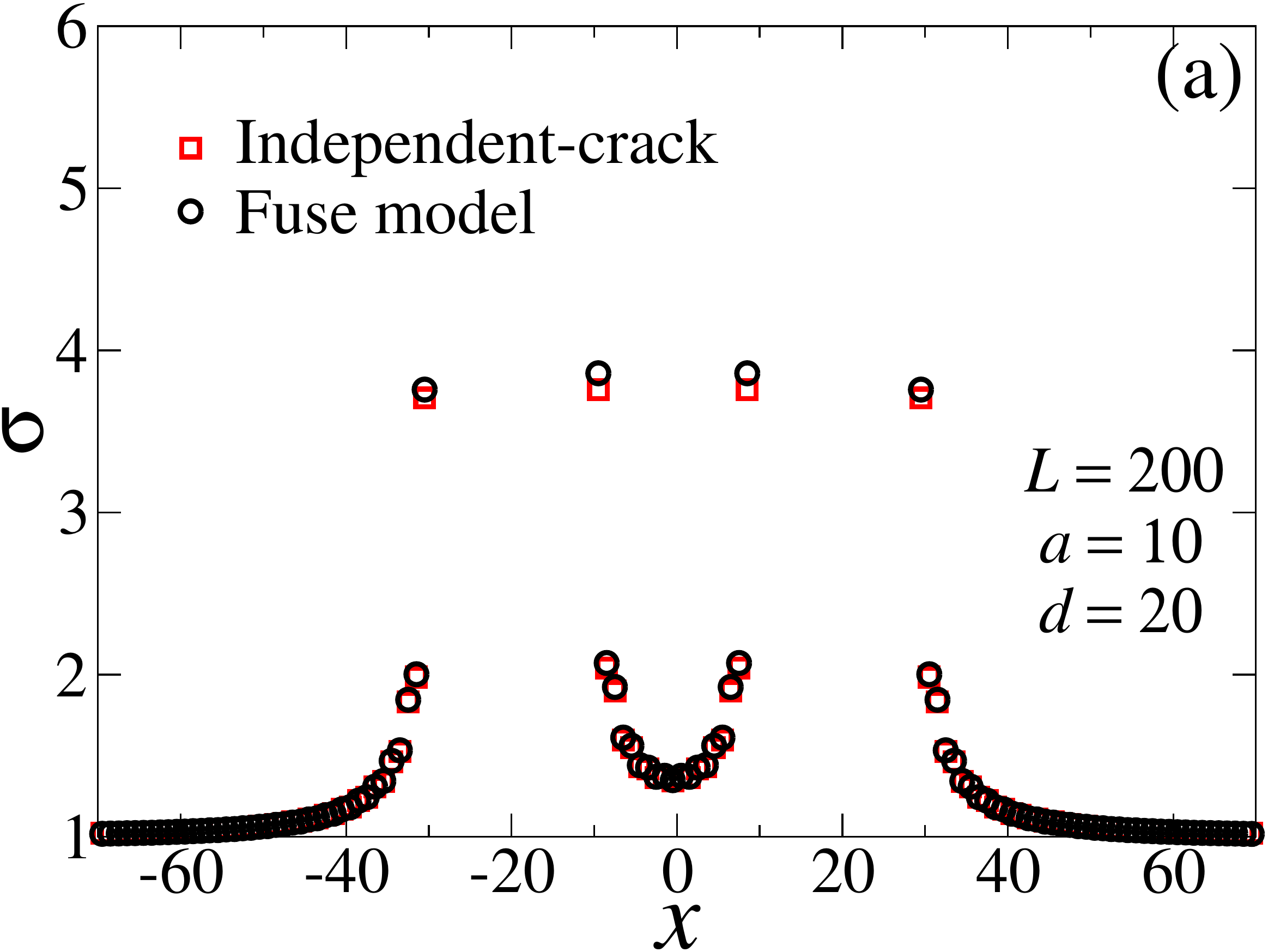} 
}\\
\subfloat{
\includegraphics[width=\columnwidth]{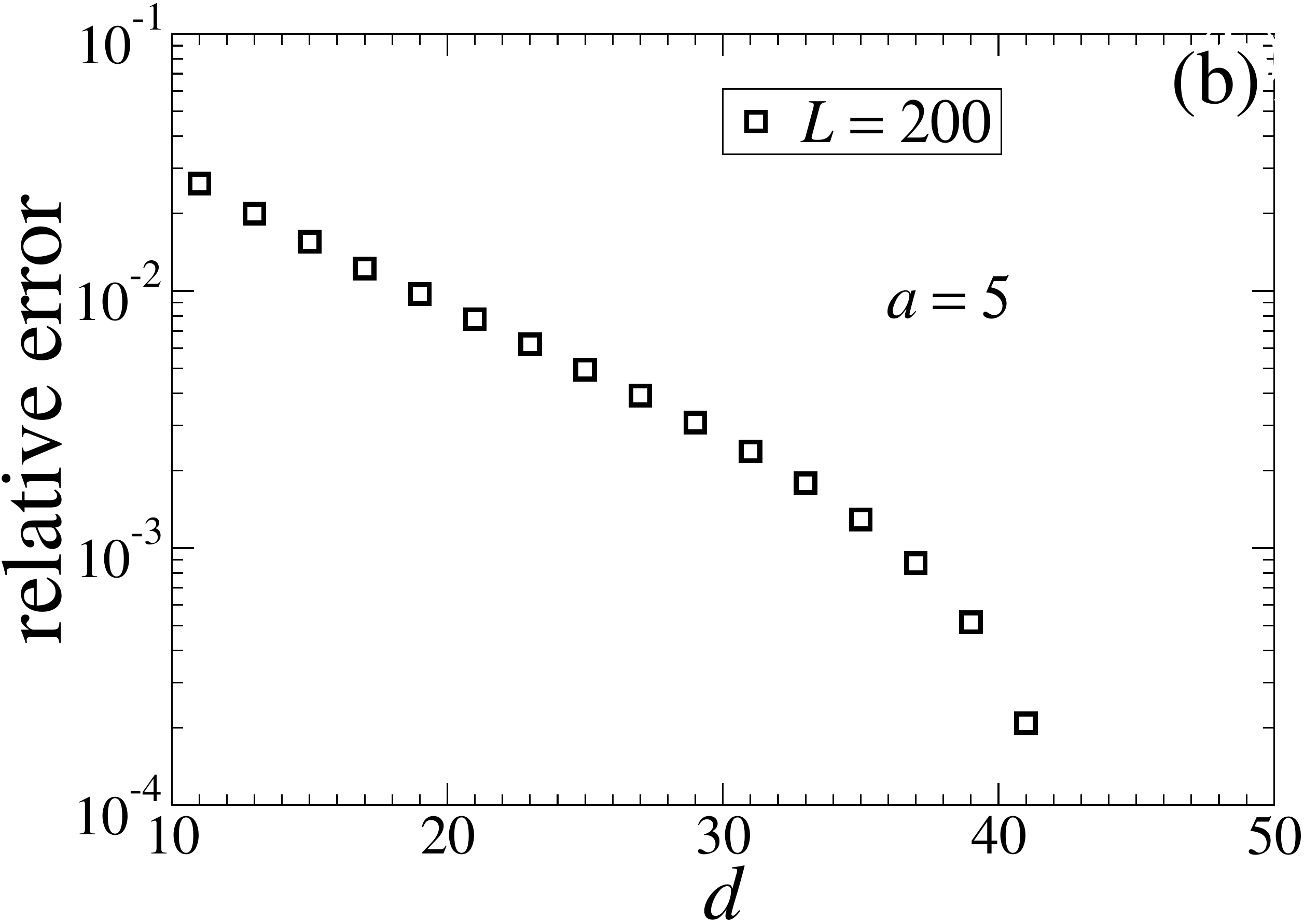}
}
\caption{(Color online) Top: comparison between the stress along the propagation 
line of the system calculated exactly (black circles) and by the independent-crack
approximation (red squares) for a sample of length $L$ containing two cracks 
of length $2a$ separated by a distance $d$. 
Both calculations were performed for the fuse model,
which is equivalent to fracturing a discretized scalar linear-elastic theory (see main text).
The independent-crack approximation uses the stress field calculated within the fuse
model as if each crack would be separately present in the system.
Bottom: relative error between the exact result and the independent-crack
approximation for the stress at the crack tip, as a function
of the separation $d$ between cracks of length $2a$.}
\label{fig14}
\end{figure}

Our approximate results can be compared with another approach, the
\emph{fuse model} \cite{Gilabert1987,Arcangelis1985}, which is equivalent
to fracturing a discretized scalar version of linear-elastic theory, appropriate for the loading mode 
and the two-dimensional geometry we assume here. Within the fuse model, we can compute numerically the
finite-size value of the local stress in multicrack configurations.

The independent-crack approximation (ICA) consists in writing the stress 
(and thus also the stress amplitude)
in the element located at position $x$ when the multicrack configuration is 
$\{a_k,\overline{x}_k\}_n$
as
\begin{gather}
\sigma(x;\{a_k,\overline{x}_k\}_n) \simeq \sigma_0 +
\sum_{k=1}^{N} \left[\sigma_1\left(x;\overline{x}_k,a_k\right) - \sigma_0\right]
%
\label{independent-crack}\\
x\notin\bigcup_{k=1}^{N}\left(\overline{x}_k-a_k,\overline{x}_k+a_k\right),
\nonumber
\end{gather}
in which $\sigma_0$ is the applied external stress, $a_k$ is the
half-length of the $k$th crack, which is centered at position $\overline{x}_k$ 
with respect to the midpoint of the initial crack (which we assume again to 
be located at the center of the system), and 
$\sigma_1\left(x;\overline{x}_k,a_k\right)$ is the stress field which would be produced
by the $k$th crack in case it were the only crack in the system. 
The $-\sigma_0$ factors inside the square brackets
on the right-hand side of Eq. 
\ref{independent-crack} ensure that very far from any cracks the external
stress is recovered. Inside any of the cracks, the stress is zero.

In order to get an idea about the accuracy of the ICA, we compare its predictions
with those of the fuse model for the case in which there are two symmetric 
cracks with length $2a$ whose centers are separated by $d$ elements. The numerical
comparison is shown in Fig. \ref{fig14}, and indicates good qualitative and quantitative
agreement, with a relative error of at most a few percent.

We now discuss the results obtained by implementing the disordered crack-growth model 
according to the ICA with $\gamma>0$, presenting comparisons with the 
random fuse model whenever appropriate. In our simulations we performed averages over
up to $100\,000$ disorder realizations, with system sizes ranging from 
$L=2^5$ to $L=2^9$. We varied the damage-accumulation exponent 
$\gamma$ and the disorder strength $\Delta F$. The single-crack stress fields
$\sigma_1\left(x;\overline{x}_k,a_k\right)$ were calculated from Eq. (\ref{stress-range}).

First we note that it can be shown (see Ref. \cite{Vieira2008}) that for $\gamma<2$
any amount of disorder leads to the appearance of secondary cracks, while for 
$\gamma>2$ those appear only for stronger disorder, such that $F_1/F_2\lesssim
1 - 1/\zeta\left(\frac{1}{2}\gamma\right)$, which, in terms of the disorder strength
$\Delta F$, corresponds to
\begin{align}
 \Delta F > \Delta F_\mathrm{min} \simeq \frac{2}{2\zeta\left(\frac{1}{2}\gamma\right) - 1}.
 \label{eq:deltaFmin}
\end{align}
The value of $\Delta F_\mathrm{min}$ monotonically increases from $0$ at $\gamma=2$
to $2$ as $\gamma\rightarrow\infty$, which implies that, for large values of $\gamma$,
secondary cracks appear only if the disorder distribution allows the presence of 
arbitrarily small local damage thresholds.

\begin{figure}
\centering
\subfloat{
\includegraphics[width=\columnwidth]{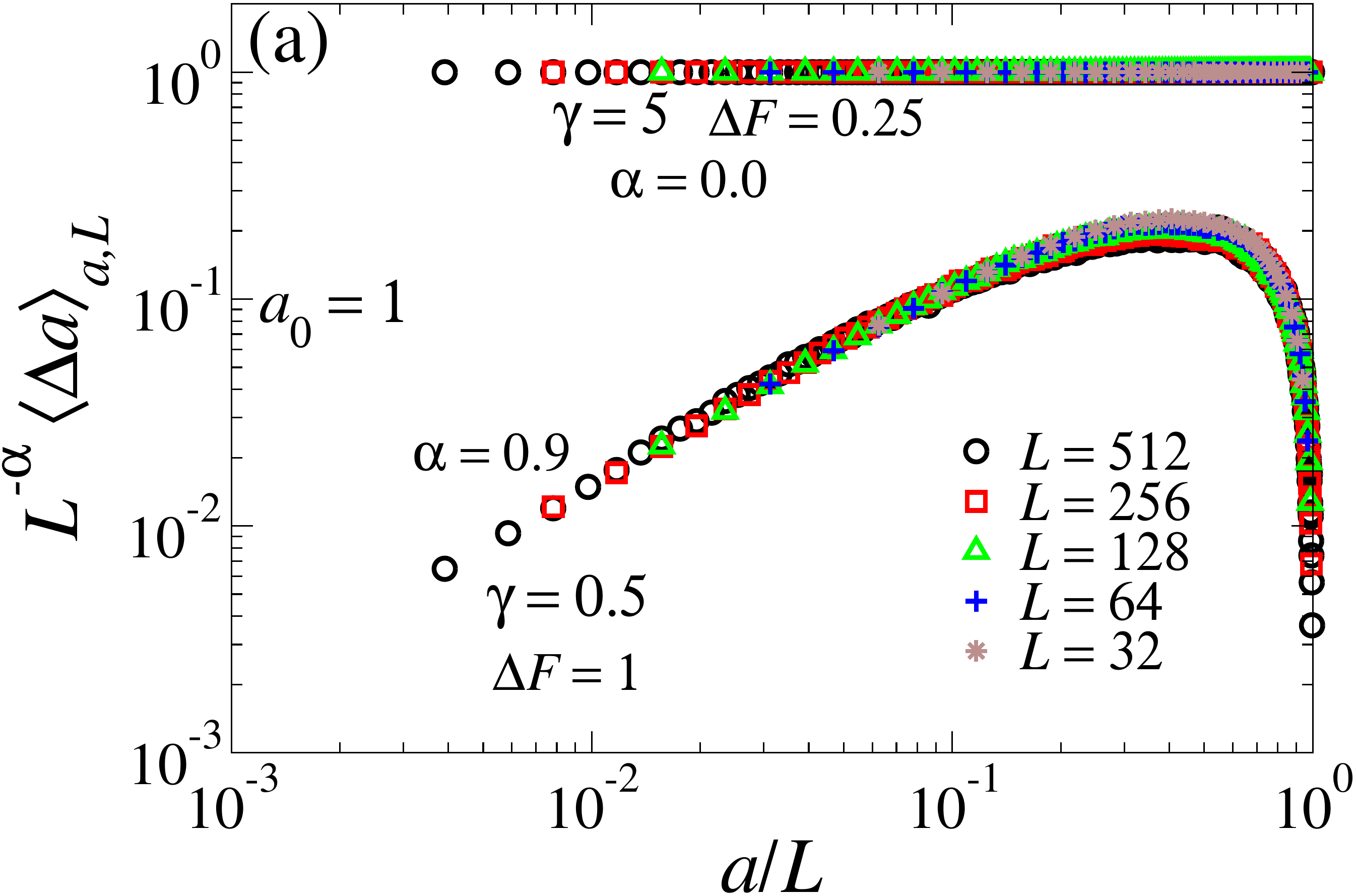} 
}\\
\subfloat{
\includegraphics[width=0.5\columnwidth]{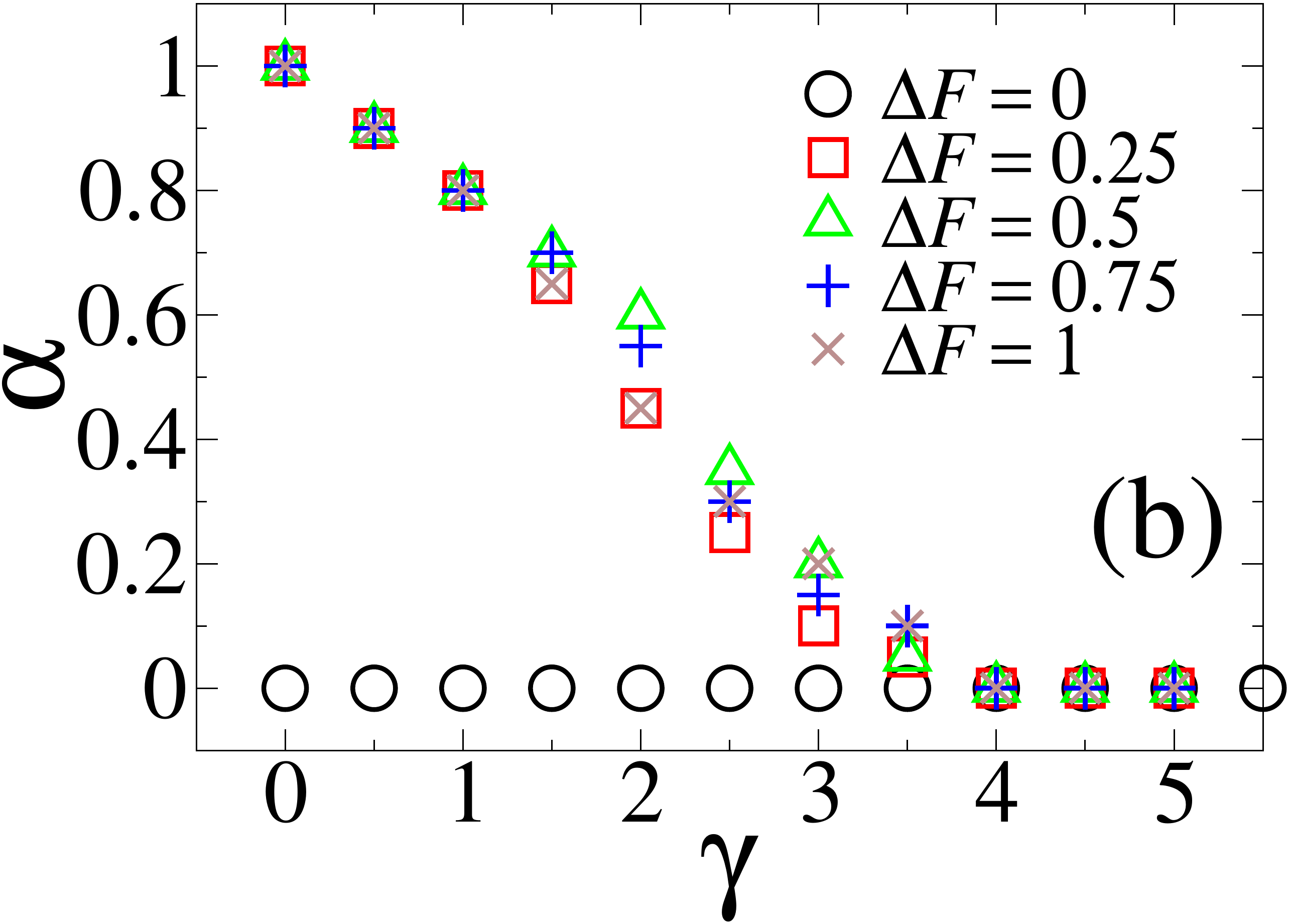}
}
\subfloat{
\includegraphics[width=0.5\columnwidth]{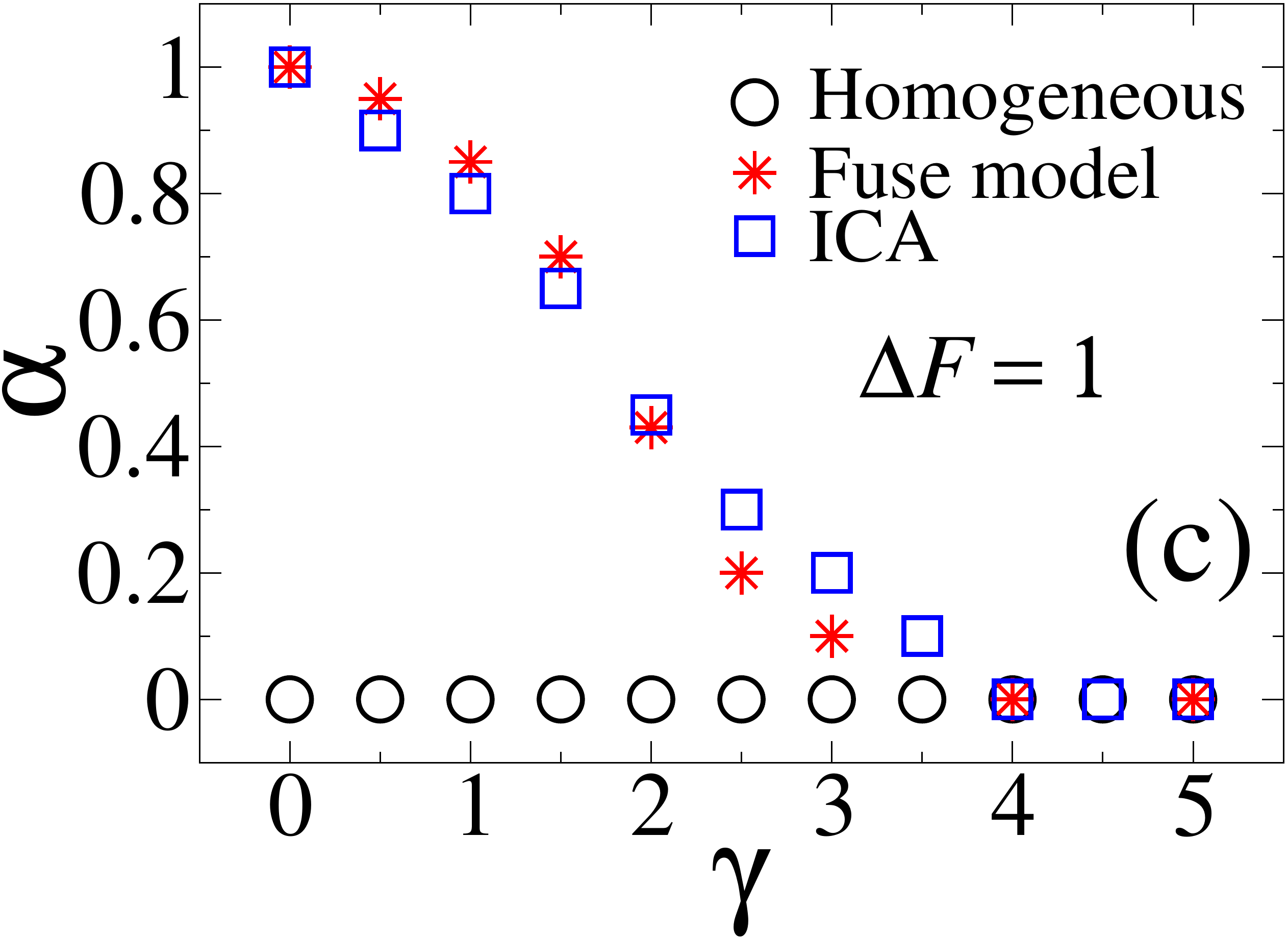}
}
\caption{(Color online) Top: scaling plot of the average main crack jump 
$\langle\Delta a\rangle_{a,L}$ as a function of the rescaled crack half-length $a/L$, 
for different sample sizes ranging from $L=2^5$ to $L=2^9$ 
and two values of the damage-accumulation exponent $\gamma$ and the disorder
strength $\Delta F$.
Bottom: dependence of 
the power-law exponent $\alpha$ on the damage accumulation 
exponent $\gamma$ for different degrees of disorder, as predicted by the ICA (left) 
and comparison between predictions of the ICA and the random fuse model
for $\Delta F=1$ (right).}
\label{fig15}
\end{figure}

\begin{figure}
\centering
\subfloat{
\includegraphics[width=\columnwidth]{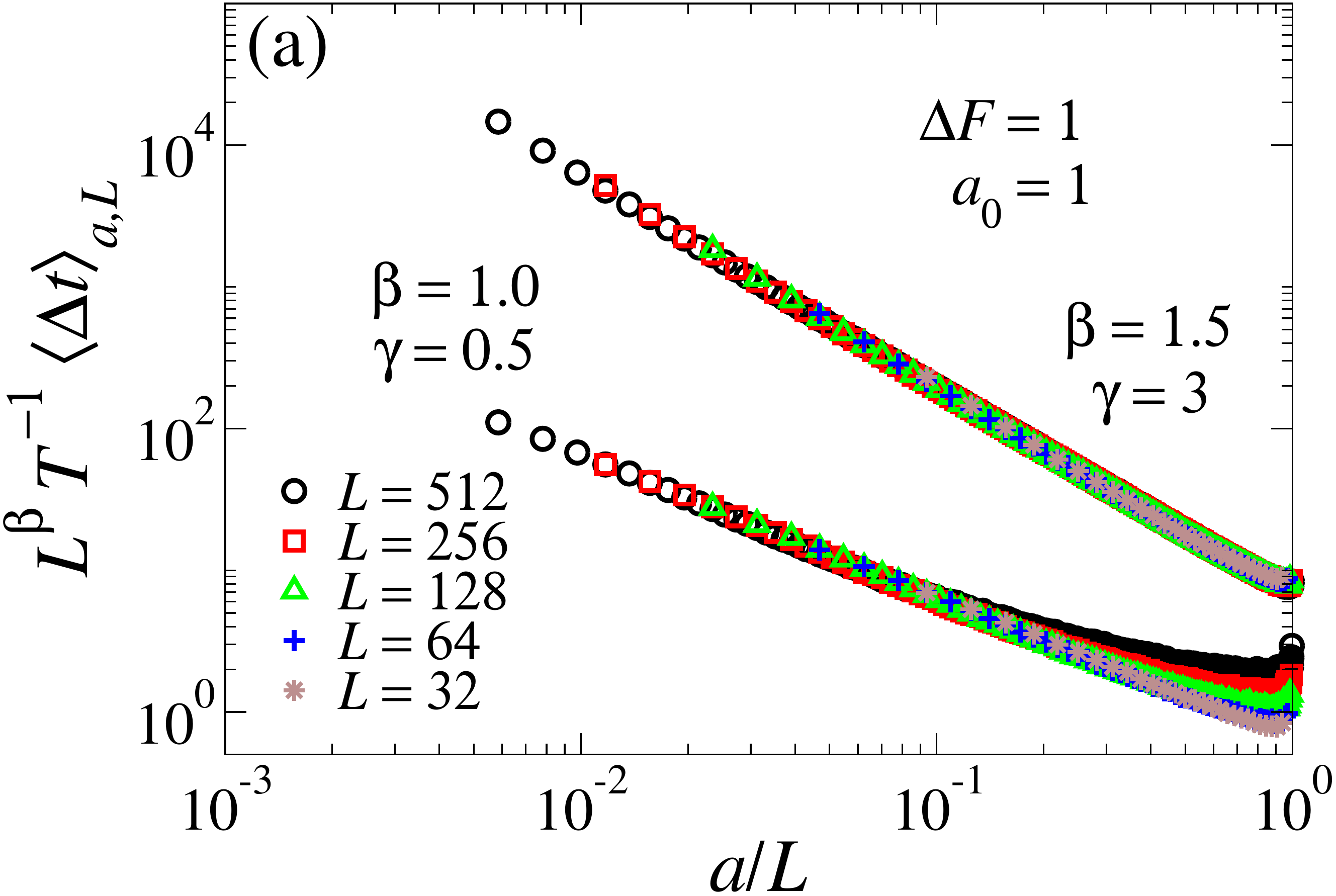} 
}\\
\subfloat{
\includegraphics[width=0.5\columnwidth]{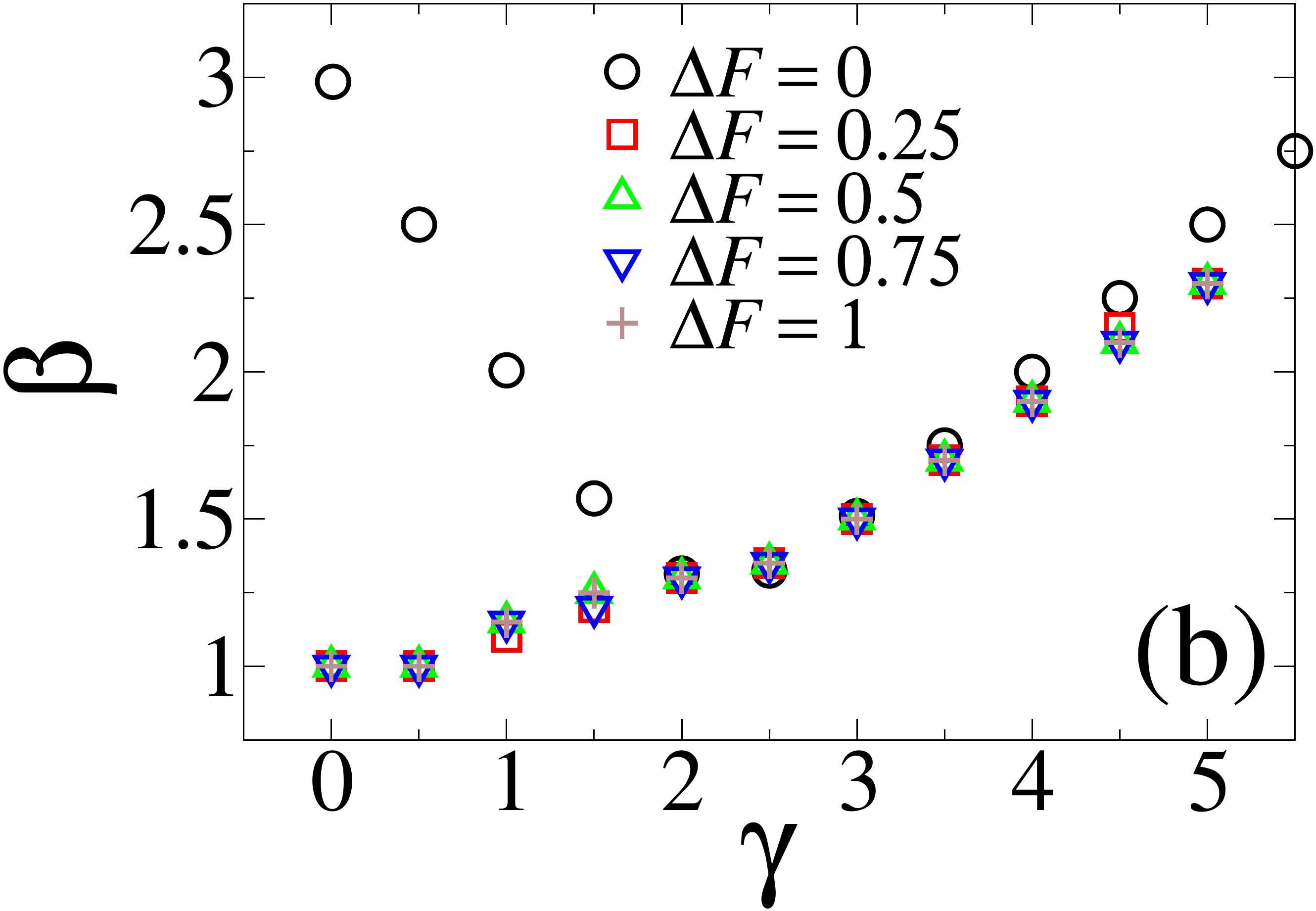}
}
\subfloat{
\includegraphics[width=0.5\columnwidth]{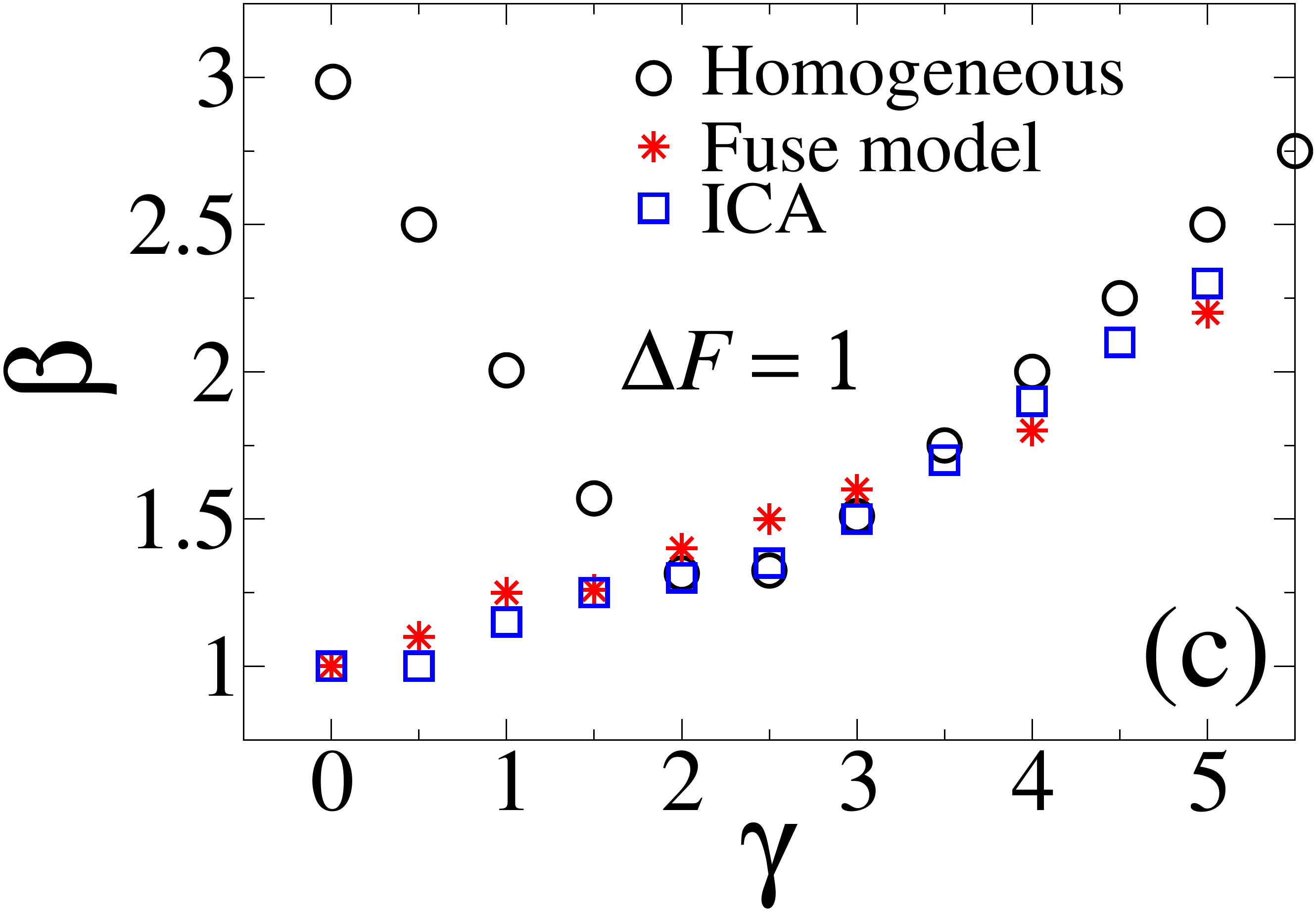}
}
\caption{(Color online) 
Top: scaling plot of the average waiting time between successive jumps
of the main crack, $\langle\Delta t\rangle_{a,L}$, normalized 
by the average rupture time $T$, as a function of the rescaled half-length $a/L$, 
for different sample sizes ranging from $L=2^5$ to $L=2^9$ 
and a few values of the damage-accumulation exponent $\gamma$ and $\Delta F=1$.
Bottom row: Dependence of 
the power-law exponent $\beta$ on the damage accumulation 
exponent $\gamma$ for different degrees of disorder, as predicted by the ICA (left) 
and comparison between predictions of the ICA and the random fuse model
for $\Delta F=1$ (right).
}
\label{fig17}
\end{figure}

For all values of $\gamma$,
both the average crack jump (avalanche size) $\Delta a$ and the average 
waiting times between consecutive jumps $\Delta t$ seem to follow 
power laws of the main crack length $2a$, namely 
$\langle\Delta a\rangle_{a,L}\sim a^\alpha$ and 
$\langle\Delta t\rangle_{a,L}\sim a^{-\beta}$,
as shown by the finite-size
scaling plots of Figs. \ref{fig15} and \ref{fig17}. 
The results for the corresponding exponents $\alpha$ and $\beta$
are in good agreement with those predicted by the random fuse model. Notice that $\alpha$
quickly becomes negligible for $\gamma>\gamma_c$, indicating that in this regime
the formation of secondary cracks is rare, except in the presence of strong
disorder ($\Delta F>\Delta F_\mathrm{min}$). As for the $\beta$
exponent, it seems to be approximately given by $\gamma/2$ for $\gamma>2$,
while approaching $\beta = 1$ as $\gamma\rightarrow 0$.

Predictions of the ICA for the average crack growth rate of the main crack
are shown in the finite-size scaling plots of Fig. \ref{fig17}, exhibiting
the power-law behavior associated with the Paris law.
The values of the Paris exponent are chosen so as to yield the best data 
collapse of the curves corresponding to different system sizes for the same 
values of the damage-accumulation exponent $\gamma$, with the 
help of Eq. (\ref{fss2}). The dependence of the 
macroscopic Paris exponent $m$ on the damage-accumulation exponent 
$\gamma$, for different degrees of disorder, is shown in Fig. \ref{fig16}, 
together with the results found for the homogeneous 
case \cite{Vieira2008} and the random fuse model \cite{Oliveira2012}.

\begin{figure}
\centering
\includegraphics[width=\columnwidth]{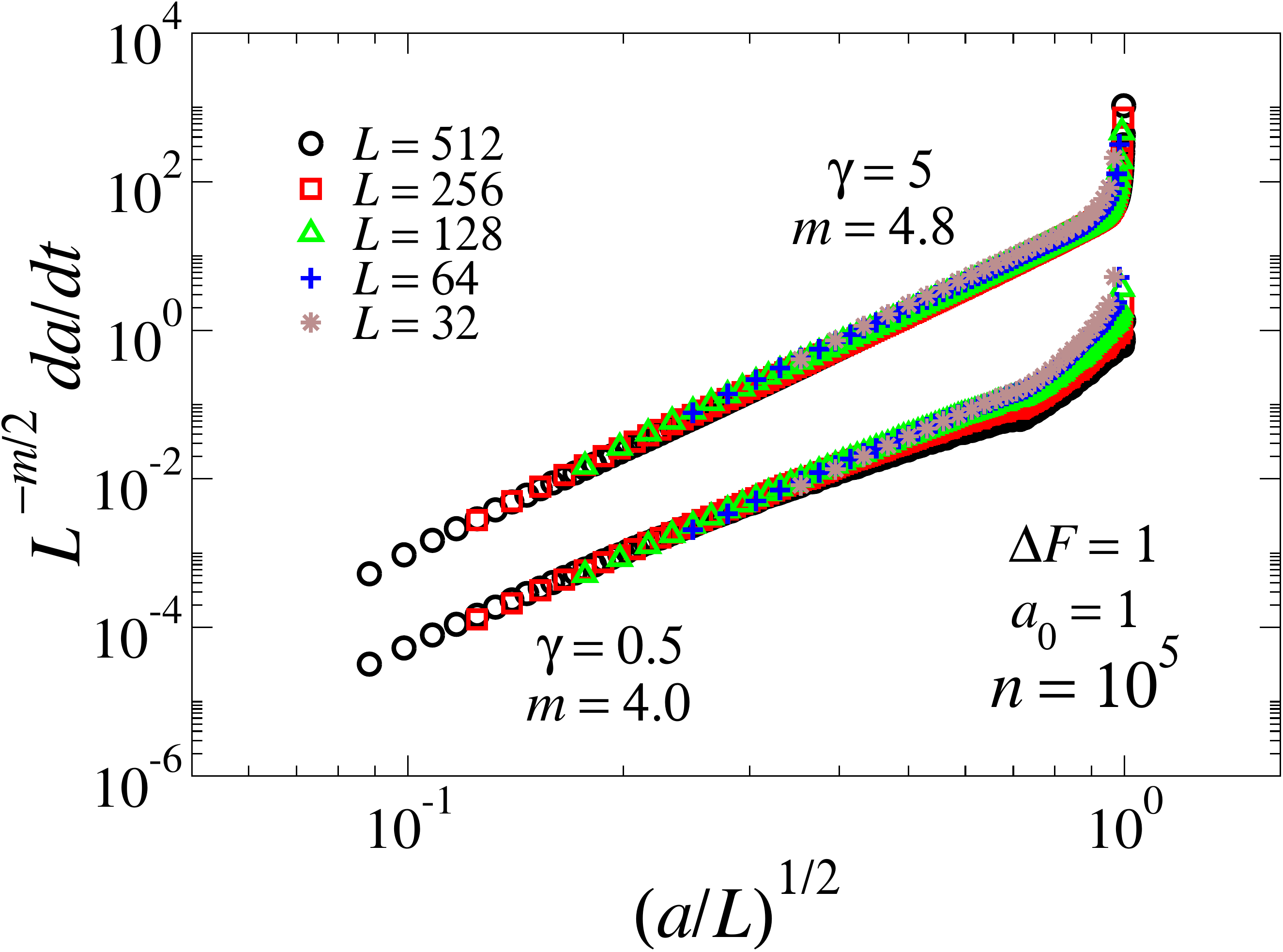}
\caption{(Color online) Scaling plot of the main crack growth rate $da/dt$ as a 
function of the crack rescaled half-length $a/L$, for different system sizes 
(from $L\,{=}\,2^5$ to $L\,{=}\,2^9$) and a few values of the damage-accumulation 
exponent $\gamma$. The disorder strength is fixed at $\Delta 
F\,{=}\,1$). Curves for $\gamma=5$ are offset for clarity.
In order to minimize statistical fluctuations, crack-growth rates 
were calculated from the numerical derivative of the half-crack length 
with respect to the average time in which the crack became trapped in a 
configuration with the corresponding length. Averages were taken over
$n\,{=}\,10^5$ disorder realizations.
}
\label{fig18}
\end{figure}

Notice that, in all the cases studied, we observed a 
strong tendency of the Paris exponent for $\gamma\lesssim 2$
to display a value $m(\gamma)\simeq 4$, irrespective of
the disorder strength. This can be understood on the basis of the 
observation that, already in the uniform limit, $\gamma_c=2$ 
separates a growth regime in which damage accumulation happens
mostly around the crack tips ($\gamma>2$) from another regime
where damage accumulation accumulates more uniformly along
the propagation line ($\gamma<2$). It is thus not surprising
that, upon the introduction of random damage thresholds,
this last regime is dominated by disorder effects, rather
than by the relatively small variations in damage accumulation
along the propagation line, therefore leading to $m=4$, as
in the $\gamma\rightarrow 0$ limit. 
On the other hand, for $\gamma\gtrsim 4$ the Paris exponent $m(\gamma)$
assumes values very close to the uniform-limit result $\gamma$,
as already observed in the random-fuse calculations
\cite{Oliveira2012}. The region $2\lesssim\gamma\lesssim 4$ is 
plagued by large statistical fluctuations and
corrections to scaling, making it difficult 
to locate within this picture.

\begin{figure}
\centering
\subfloat{
\includegraphics[width=\columnwidth]{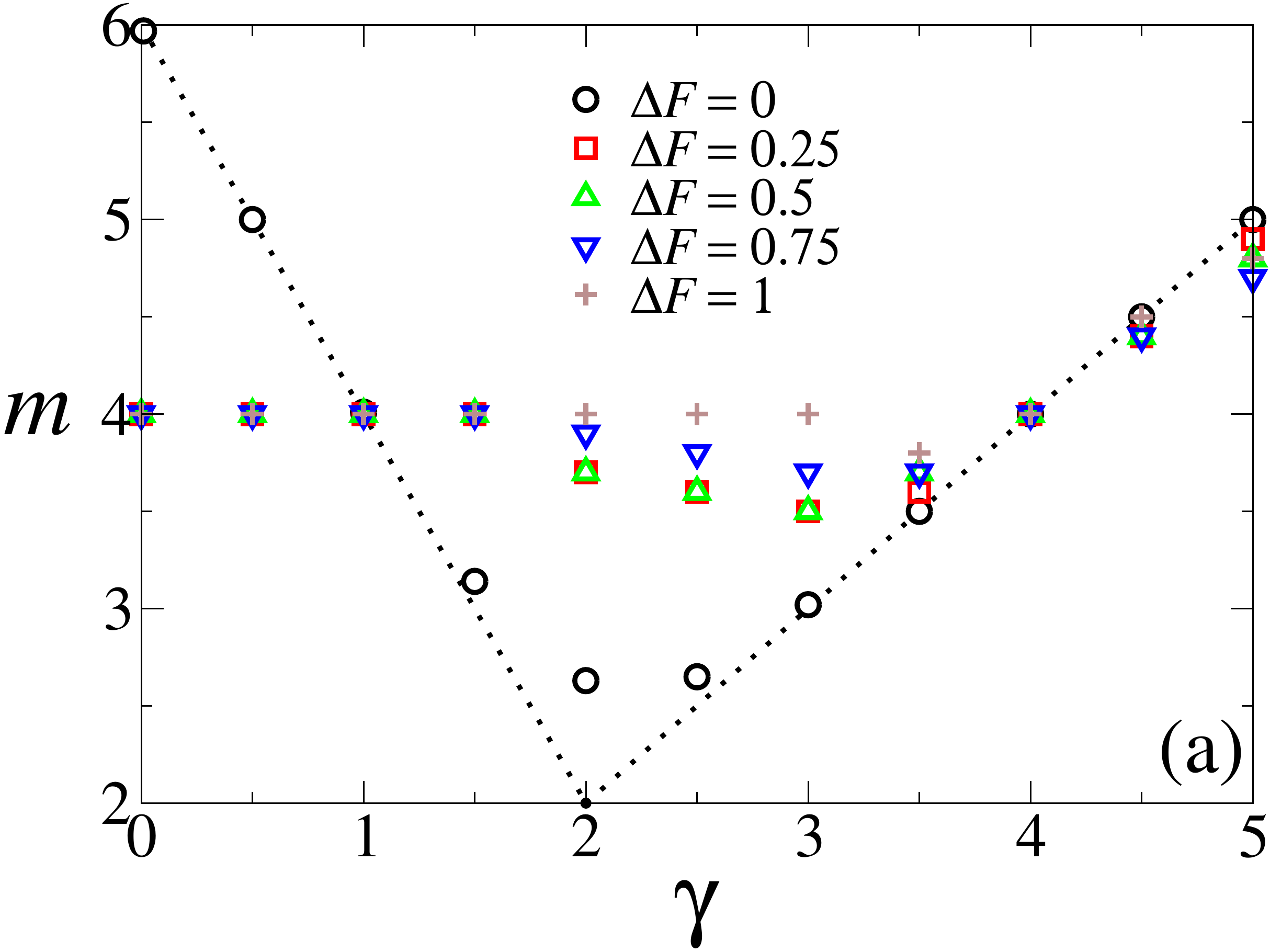} 
}\\
\subfloat{
\includegraphics[width=\columnwidth]{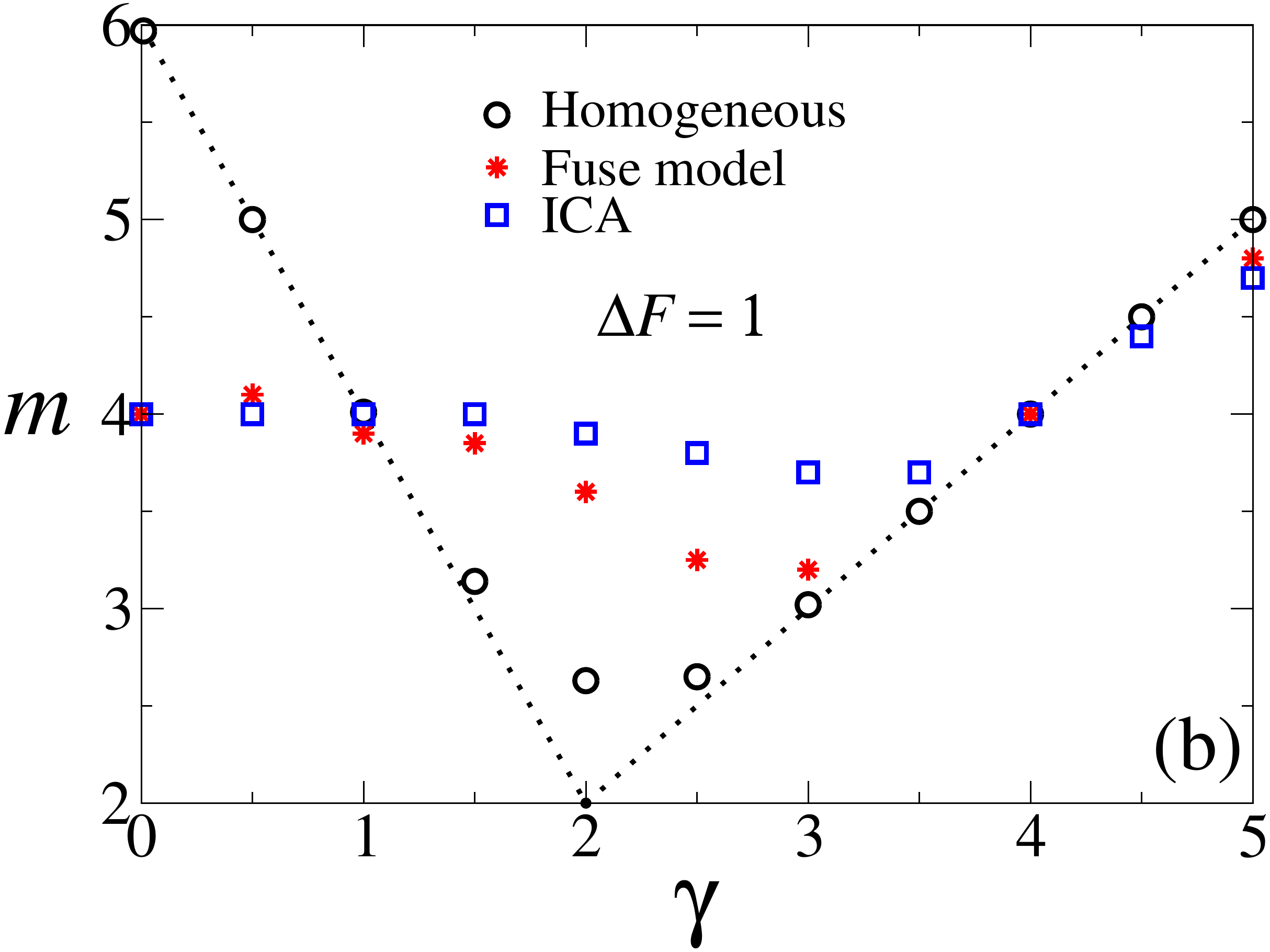}
}
\caption{(Color online) Top: dependence of the Paris exponent $m$ on
the damage-accumulation exponent $\gamma$ for the disordered model, 
according to the independent-crack approximation. 
Bottom: comparison between the results obtained by the independent-crack 
approximation and the random fuse model for the same relation 
$m\,{\times}\,\gamma$, with disorder strength $\Delta F=1$. 
Notice the good agreement except in the vicinity of $\gamma=2$.}
\label{fig16}
\end{figure}

\section{Healing effects in the uniform limit\label{sec:healing}}

We finally return briefly to the uniform limit, and introduce the
possibility of damage healing with a characteristic time
$\tau$. Explicitly, we assume that, up to time $t$,
the accumulated damage on the element
located at position $x$ is given by \cite{Kun2007}
\begin{align}
 F\left(x;t\right) = f_0 \int_0^t dt^{\prime}\left[\Delta\sigma\left(x;t^\prime\right)\right]^\gamma
 e^{-\left(t-t^\prime\right)/\tau},
\end{align}
where $f_0$ is a constant setting the time scale, $\Delta\sigma\left(x;t\right)$
is the stress amplitude at position $x$ and time $t$, and $\gamma$ is the damage
amplification exponent. Healing mechanisms during fatigue crack growth are known to be relevant, for instance,
in materials such as asphalt \cite{Zhiming2002} and also in self-healing composite materials such as epoxy, with
the incorporation of microencapsulated healing agents such as dicyclopentadiene \cite{Brown2005}.
The healing time $\tau$ is treated here as another
phenomenological parameter, which presumably depends on the temperature and possibly
on the concentration of a healing agent.

Taking into account that $\Delta\sigma\left(x;t\right)$ does not vary between 
crack growth events, the last equation leads to a recursion relation
for the damage at a given location when the crack has length $2a$,
\begin{align}
 F(x;a) = e^{-\delta t(a)/\tau}F(x;a-\delta r) + \delta F(x;a),
\end{align}
with
\begin{align}
 \delta F(x;a)=f_0\tau\left[\Delta\sigma(x;a)\right]^\gamma\left(1-e^{-\delta t(a)/\tau}\right),
\end{align}
where the symbols have the same meaning as in Sec. \ref{model}, and we have
used the fact that in the uniform limit the crack always grows by the breaking
of the elements at the crack tips. Notice that as $\tau\rightarrow\infty$ we
recover Eqs. (\ref{damage}) and (\ref{recurrence}).

The time interval $\delta t(a)$ during
which the crack has length $2a$ is determined from the condition $F(a+\delta r;a)=
F_\mathrm{thr}$. For the time during which the crack remains with the initial notch
size $2a_0$ this yields
\begin{align}
 \delta t(a_0) = -\tau\ln\left(1-\frac{F_\mathrm{thr}}
 {f_0\tau\left[\Delta\sigma(a_0+\delta r;a_0)\right]^\gamma}\right),
\end{align}
indicating the existence of a minimum value of $\tau$ below which the crack
cannot grow. This minimum value is given by
\begin{align}
 \tau_\mathrm{min} = \frac{F_\mathrm{thr}}
 {f_0\left[\Delta\sigma(a_0+\delta r;a_0)\right]^\gamma}.
\end{align}
For a fixed value of $\tau$, this result is compatible with the existence of a
minimum stress amplitude around which the fatigue lifetime diverges \cite{Brown2005}.

\begin{figure}
\centering
\subfloat{
\includegraphics[width=0.8\columnwidth]{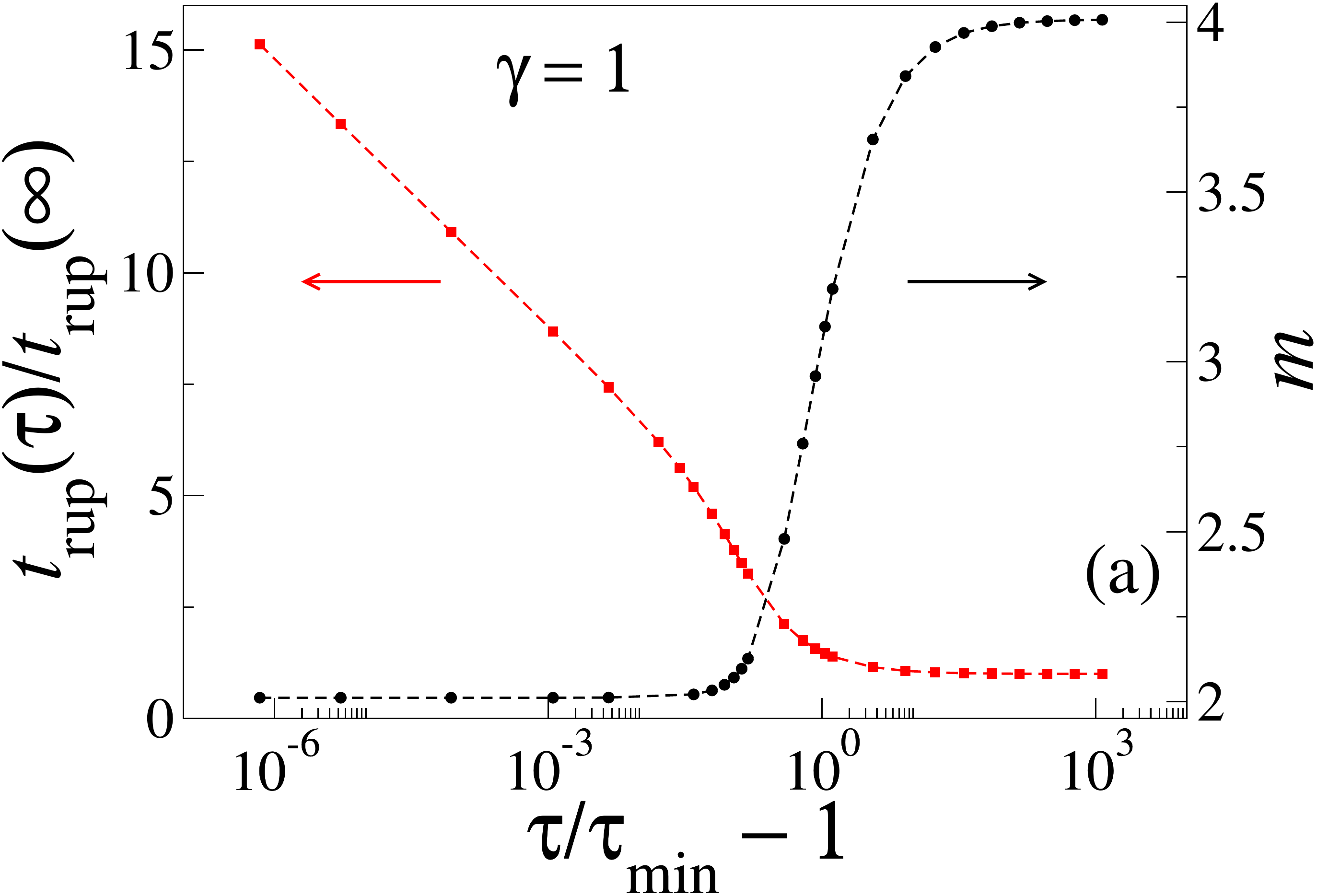} 
}\\
\subfloat{
\includegraphics[width=0.8\columnwidth]{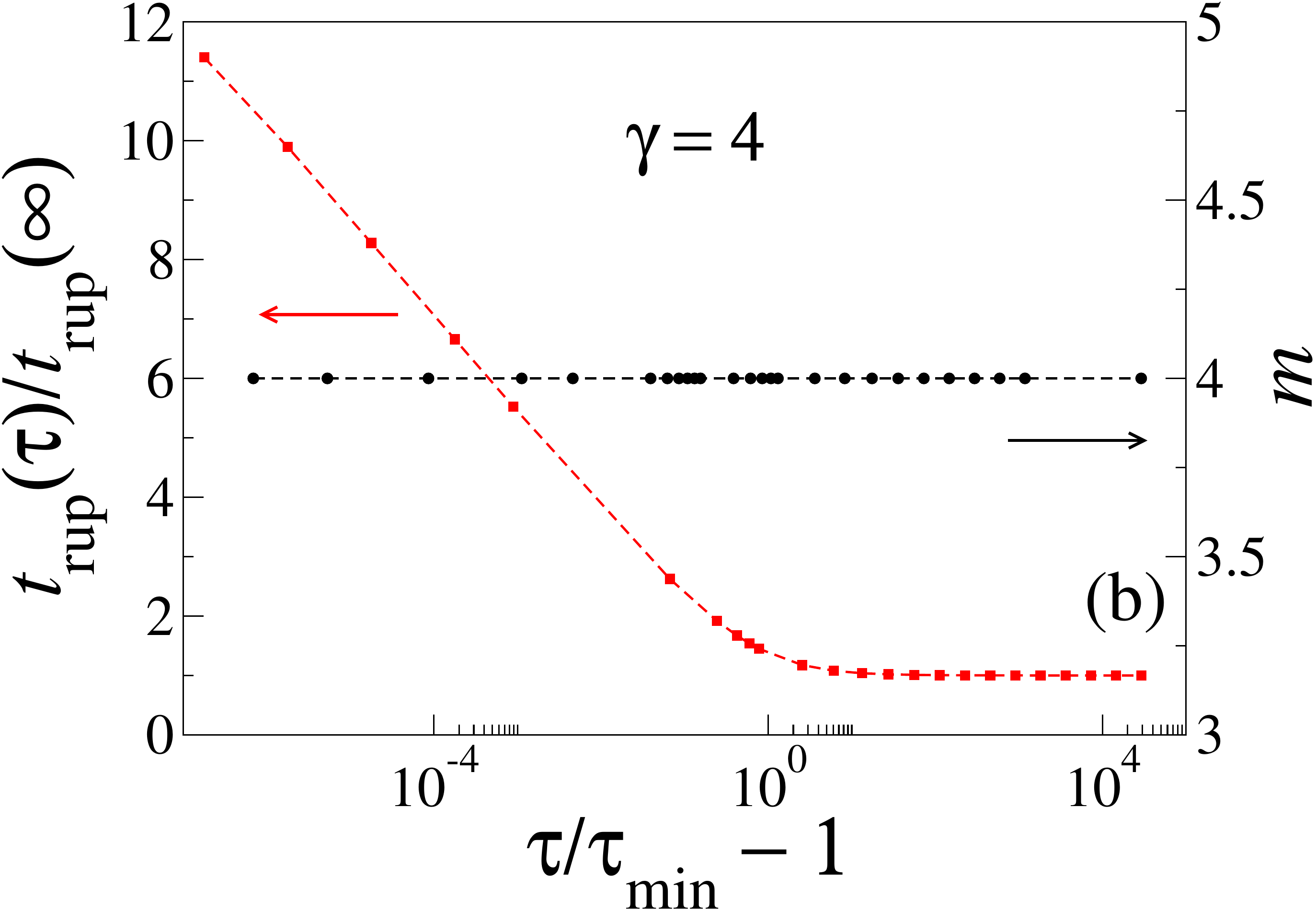}
}
\caption{(Color online) Behavior of the rescaled rupture time (red curves)
and the Paris exponents (black curves) as functions of the healing characteristic
time $\tau$, rescaled by the corresponding minimum value, for $\gamma=1$ (top)
and $\gamma=4$ (bottom).}
\label{fig:heal1}
\end{figure}

Using the previous equations we can numerically investigate the crack growth
dynamics and its dependence on the parameters $\gamma$ and $\tau$. It turns
out that the Paris exponent $m$ is independent of $\tau$ for $\gamma\geq 2$, but
becomes $\tau$-dependent for $\gamma<2$. In this last regime, $m$ is equal
to $6-2\gamma$ for $\tau\rightarrow\infty$, but it approaches the value $2$ as
$\tau$ approaches $\tau_\mathrm{min}$. Figure \ref{fig:heal1}
shows, for $\gamma=1$
and $\gamma=4$, the behavior of $m$ as a function of $\tau$ for a finite sample
with $L=2^{15}$ elements. Also shown is the $\tau$ dependence of the rupture time
$t_\mathrm{rup}$, normalized by its value in the limit $\tau\rightarrow\infty$.
Notice the seemingly logarithmic divergence of $t_\mathrm{rup}$ as $\tau
\rightarrow\tau_\mathrm{min}$, a prediction whose experimental verification
would require an estimate of the healing time $\tau$ in terms of material and 
environmental parameters. At the moment, to the best of our knowledge, such
estimates are not available.

\section{Conclusions}

In summary, we investigated various extensions of a 
model for subcritical fatigue crack growth in which 
damage accumulation is assumed to follow a power law of the local
stress amplitude. In all cases, our main interest was in determining
the effects of model ingredients on the Paris exponent
governing subcritical crack-growth
dynamics at the macroscopic scale, starting from a single small
notch propagating along a fixed line. 

In the uniform limit, we showed that a number of analytical 
and numerical results can be established regarding the dependence of 
the Paris exponent on the damage-accumulation exponent, the threshold stress
range required to induce local damage, and the characteristic
time of damage healing. There is a critical value of the damage 
accumulation exponent,
namely $\gamma_c=2$, separating two distinct regimes of behavior
for the Paris exponent $m$. For $\gamma>\gamma_c$, the Paris exponent
is shown to assume the value $m=\gamma$, a result which proves robust
against the introduction of various modifying ingredients. On the 
other hand, in the regime $\gamma<\gamma_c$ the Paris exponent
is seen to be sensitive to the different ingredients added to the model,
with rapid healing or a threshold stress amplitude $b=1$ leading
to $m=2$ for all $\gamma<\gamma_c$,
in contrast to the linear dependence $m=6-2\gamma$ observed for very
long characteristic healing times and $b=0$.

The introduction of disorder on the local fatigue thresholds
leads to the possible appearance of multiple cracks along the
propagation line, and the Paris exponent tends to $m\simeq 4$ for
$\gamma\lesssim 2$, while retaining the behavior $m=\gamma$ for $\gamma>4$.
The independent-crack approximation employed for all calculations in 
the presence of disorder yields results in good agreement with the
more computationally expensive random-fuse calculations, suggesting
that it can be reliably applied to further extensions of the model.
An interesting candidate would be an investigation of the combined
effects of disorder and healing, a situation which is closer to
what occurs in real materials.

It is possible to compare the results obtained from the present approach
with those derived in recent years (see e.g. Refs.
\cite{Ciavarella2008,Paggi2009,Carpinteri2011,Paggi2014,Jones2016})
based on the extension of ideas
of incomplete self-similarity as applied directly to the macroscopic
Paris law (see e.g. Refs. \cite{Ritchie2005,Barenblatt2006} and references
therein). These works point not
only to the effect, on the Paris exponent,
of characteristic lengths (usually the sample
thickness) or of plasticity properties of the fracture-process zone ahead 
of the crack tip \cite{Ritchie2005},
but also to the fact that
the fractal character of the crack profile leads to modifications
of the asymptotic behavior of the stress field around the crack tip,
which also affects the Paris law.
Specifically, this changes the dependence of the stress field on the 
distance $r$ to a thin crack tip, which now diverges as $r^{(D-2)/2}$,
$D$ being the fractal dimension of the crack profile
\cite{Yavari2002}. Notice that
this makes the stress field decay more slowly with $r$ than the $r^{-1/2}$
behavior of a linear ($D=1$) crack. This is reminiscent of the behavior
of a damage-accumulation rule with $\gamma<2$, for which, as discussed
in Sec. \ref{sec:uniform}, damage is more uniformly distributed along
the crack line. Therefore, a possible interpretation of the present
approach is that, via the introduction of the damage-accumulation 
exponent $\gamma$, it encapsulates various effects such as the plasticity properties
ahead of the crack tip and the fractal nature of the crack profile,
allowing the use of linear-elastic fracture mechanics to provide an effective
description of fatigue crack dynamics.

Incidentally, the question remains as to whether it is possible
to relate the phenomenological, mesoscopic damage-accumulation
exponent $\gamma$ to atomistic or structural 
features of real materials. We are currently investigating the 
possibility of employing molecular dynamics or phase-field
methods to approach this issue.


\begin{acknowledgments}
We thank the Brazilian 
agencies FAPESP and CNPq for their financial support. 
MSA thanks Carmen Prado and André Timpanaro for useful 
discussions.
We acknowledge financial support from the European Research Council (ERC) 
Advanced Grant 319968-FlowCCS and from NAP-FCx. 
\end{acknowledgments}

\bibliographystyle{apsrev4-1}
\bibliography{referencias-doutorado}

\end{document}